\documentclass[preprint,3p]{elsarticle}

\usepackage[%
                  bookmarks,%
                  bookmarksopen=false,%
                  pdfauthor={Autor},%
                  pdftitle={Titel},%
                  colorlinks=true,%
                  linkcolor=blue,%
                  citecolor=red,%
                  urlcolor=green,%
                ]{hyperref}
\usepackage{aliascnt}

\biboptions{authoryear}

\usepackage{natbib}

\usepackage{amssymb}
\usepackage{amsfonts}
\usepackage{amsmath}
\usepackage{amsthm}
\usepackage{graphicx}
\usepackage{multicol}
\usepackage{multirow}
\usepackage{caption}
\usepackage{subcaption}
\usepackage[usenames, dvipsnames]{xcolor}
\usepackage{verbatim}
\usepackage{color}
\usepackage[numbers]{natbib}
\usepackage{relsize}
\usepackage{lmodern}
\usepackage{url}
\usepackage{MnSymbol}
\usepackage{tikz}
\usepackage{float}
\usepackage{multicol}
\usepackage{vwcol}
\usetikzlibrary{arrows}

\theoremstyle{plain}
\newtheorem{theorem}{Theorem}[section]
\newaliascnt{lemma}{theorem}
\newtheorem{lemma}[lemma]{Lemma}
\aliascntresetthe{lemma}

\newaliascnt{remark}{theorem}
\newtheorem{remark}[remark]{Remark}
\aliascntresetthe{remark}

\newaliascnt{proposition}{theorem}
\newtheorem{proposition}[proposition]{Proposition}
\aliascntresetthe{proposition}

\newaliascnt{definition}{theorem}
\newtheorem{definition}[definition]{Definition}
\aliascntresetthe{definition}

\newaliascnt{corollary}{theorem}
\newtheorem{corollary}[corollary]{Corollary}
\aliascntresetthe{corollary}

\newaliascnt{example}{theorem}
\newtheorem{example}[example]{Example}
\aliascntresetthe{example}

\newaliascnt{algorithm}{theorem}
\newtheorem{algorithm}[algorithm]{Algorithm}
\aliascntresetthe{algorithm}

\newtheorem{fig}[figure]{Figure}

\newcommand{\balg}{\begin{algorithm}\rm}
\newcommand{\ealg}{\end{algorithm}}

\newcommand{\bthe}{\begin{theorem}}
\newcommand{\ethe}{\end{theorem}}

\newcommand{\ben}{\begin{enumerate}}
\newcommand{\een}{\end{enumerate}}

\newcommand{\bit}{\begin{itemize}}
\newcommand{\eit}{\end{itemize}}

\newcommand{\beq}{\begin{equation}}
\newcommand{\eeq}{\end{equation}}

\newcommand{\ble}{\begin{lemma}}
\newcommand{\ele}{\end{lemma}}

\newcommand{\bde}{\begin{definition}\rm}
\newcommand{\ede}{\halmos\end{definition}}

\newcommand{\bco}{\begin{corollary}}
\newcommand{\eco}{\end{corollary}}

\newcommand{\bpr}{\begin{proposition}}
\newcommand{\epr}{\end{proposition}}

\newcommand{\brem}{\begin{remark}\rm}
\newcommand{\erem}{\halmos\end{remark}}

\newcommand{\bproof}{\begin{proof}}
\newcommand{\eproof}{\end{proof}}

\newcommand{\bexam}{\begin{example}\rm}
\newcommand{\eexam}{\halmos\end{example}}

\newcommand{\bexamwh}{\begin{example}\rm}
\newcommand{\eexamwh}{\end{example}}

\newcommand{\bfi}{\begin{fig}}
\newcommand{\efi}{\end{fig}}

\newcommand{\btab}{\begin{tab}}
\newcommand{\etab}{\end{tab}}

\newcommand{\beao}{\begin{eqnarray*}}
\newcommand{\eeao}{\end{eqnarray*}\noindent}

\newcommand{\beam}{\begin{eqnarray}}
\newcommand{\eeam}{\end{eqnarray}\noindent}

\newcommand{\ovr}{\begin{array}}
\newcommand{\barr}{\begin{array}}
\newcommand{\earr}{\end{array}}

\newcommand{\bdis}{\begin{displaymath}}
\newcommand{\edis}{\end{displaymath}\noindent}

\def\B{{\overline B}}

\def\P{{\mathbb P}}

\def\R{{\mathbb R}}

\def\cals_+{{\cals_+}}

\def\cals{{\mathcal{S}}}

\def \b {{\overline{b}}}

\newcommand{\bs}{\boldsymbol}

\newcommand{\bfx}{\boldsymbol{X}}
\newcommand{\bfsx}{\boldsymbol{x}}

\def\1{\mathds{1}}

\newcommand{\std}{\stackrel{d}{\rightarrow}}

\newcommand{\stv}{\stackrel{v}{\rightarrow}}

\newcommand{\eqd}{\stackrel{d}{=}}

\newcommand{\al}{\alpha}

\newcommand{\DAG}{\text{DAG}}
\newcommand{\TDM}{\text{TDM}}
\newcommand{\TDC}{\text{TDC}}

\newcommand{\mSEM}{\text{RMLM}}
\newcommand{\MLCM}{\text{MLCM}}

\newcommand{\ML}{\text{ML}}
\newcommand{\MDA}{\text{MDA}}
\newcommand{\RMWM}{\text{RMWM}}
\newcommand{\MLC}{\text{MLC}}

\newcommand{\sgn}{{\rm sgn}}

\newcommand{\an}{{\rm an}}
\newcommand{\pa}{{\rm pa}}

\newcommand{\des}{{\rm de}}
\newcommand{\An}{{\rm An}}
\newcommand{\Pa}{{\rm Pa}}

\newcommand{\Des}{{\rm De}}
\newcommand{\tr}{{\rm tr}}

\newcommand{\ov}{\overline}

\newcommand{\wt}{\widetilde}

\newcommand{\halmos}{\quad\hfill\mbox{$\Box$}}

\def\D{\mathcal{D}}

\definecolor{plum}{cmyk}{0.50,1,0,0}
\definecolor{TealBlue}{cmyk}{0.86,0,0.34,0.02}
\definecolor{OliveGreen}{cmyk}{0.64,0,0.95,0.40}

\usepackage{enumitem} 
\newlist{citemize}{itemize}{4} 
\setlist[citemize]{label=-,nosep,topsep=-\parskip} 

\begin{document}
\begin{frontmatter}

\title{Tail dependence of recursive max-linear models \\ with regularly varying noise variables}
\author{Nadine Gissibl}
  \ead{n.gissibl@tum.de}
\author{Claudia Kl\"uppelberg}
   \ead{cklu@tum.de}
\author{Moritz Otto}
    \ead{moritz.otto@kit.edu}

\address{Center for Mathematical Sciences, Technical University of Munich, Boltzmannstrasse 3, Garching 85748, Germany}

\begin{abstract}  
Recursive max-linear structural equation  models with regularly varying noise variables are considered. Their causal structure is represented by a directed acyclic graph (\DAG). The  problem of identifying  a recursive max-linear model and its associated \DAG\ from its matrix of pairwise tail dependence coefficients is discussed. For example, it is shown that if a causal ordering of the associated \DAG\ is additionally known, then the minimum \DAG\ representing the recursive structural equations can be recovered from the tail dependence matrix. For the relevant subclass of recursive max-linear models, identifiability of the associated minimum \DAG\ from  the tail dependence matrix and the initial nodes is shown. Algorithms find the associated minimum \DAG\ for the different situations. Furthermore, given a tail dependence matrix, an algorithm outputs all compatible recursive max-linear models and their associated minimum \DAG s.
\end{abstract}

\end{frontmatter}

\date{}

\section{Introduction} 
Causal inference is fundamental in virtually all areas of science. Examples for concepts established over the last years to understand causal inference include structural equation modeling \citep[see e.g.][]{Bollen,Pearl2009} and graphical modeling \citep[see e.g.][]{Lauritzen1996,SGS,KF}. 

In extreme risk analysis it is especially important to understand causal dependencies.   We consider {\em recursive max-linear models} (\mSEM s), which are  {max-linear  structural equation models} whose {causal structure} is represented by a {\em directed acyclic graph} (\DAG). Such models are {directed graphical models} \citep[][Theorem~1.4.1]{Pearl2009}; i.e., 
the \DAG\ encodes conditional independence relations in the distribution
 via the {(directed global) Markov property}. \mSEM s were introduced and studied in \cite{GK1}. They may find their application in situations when extreme risks play an essential role and may propagate through a network, for example, when modeling water levels or pollution concentrations in a river or when modeling risks in a large industrial structure. In \cite{einmahl2016} a \mSEM\ was fitted to data from the EURO STOXX 50 Index, where the \DAG\ structure was assumed to be known. 

In this paper we assume {\em regularly varying} noise variables. This leads to models treated in classical {multivariate extreme value theory}. The books by \cite{Beirlant2004}, \cite{DHF},  and \cite{Resnick1987,Resnick2007} provide a detailed introduction into this field.  
A \mSEM\ with regularly varying noise variables is in the {\em maximum domain of attraction} of an {\em extreme value}  ({\em max-stable}) {\em distribution}. The {spectral measure} of the limit distribution, which describes the dependence structure given by the \DAG,  is discrete. Every max-stable random vector with discrete spectral measure is {\em max-linear} (\ML), and every multivariate max-stable distribution can be approximated arbitrarily well via a \ML\ model \citep[e.g.][Section~2.2]{Yuen2014}. This demonstrates the important role of \ML\ models in extreme value theory. They have been investigated, generalized, and applied to real world problems by many researchers; see e.g. \cite{Schlather2002}, \cite{Wang2011}, \cite{FHZ}, \cite{SS}, \cite{einmahl2012m}, \cite{CZ}, and \cite{kiri2017}. 

One main research problem that is addressed for {restricted} 
recursive structural equation models, where the functions are required to belong to a specified function class, is the {\em identifiability} of the coefficients and the \DAG\  from the observational distribution. Recently, particular attention in this context has been given to recursive structural equation models with additive Gaussian noise; see e.g.  \cite{Peters2014b}, \cite{ernest2016causal}, and references therein.   For \mSEM s this problem is investigated in \citet{GKL}. In the present paper we discuss the identifiability of   \mSEM s  from their {\em (upper) tail dependence coefficients} (\TDC s). 

The \TDC, which goes back to  \cite{Sibuya1960},  measures the extremal dependence between two random variables and is a simple and popular  dependence measure in extreme value theory. Methods to construct multivariate max-stable distributions with given \TDC s have been proposed, for example, by \cite{Schlather2002}, \cite{FalkTDPM}, \cite{FHZ}, and \cite{SS}. Somehow related we identify all \mSEM s with the same given \TDC s.  

\subsection{Problem description and important concepts\label{intro3}}

First we briefly review \mSEM s  and introduce the \TDC\ formally.   We then describe the idea of this work in more detail and state the main results.  

\subsection*{Max-linear models on \DAG s}

Consider a {\em\mSEM\ $\bfx=(X_1,\ldots,X_d$)  on a \DAG\ $\D=(V,E)$} with nodes $V=\{1,\ldots,d\}$ and edges $E=\{(k,i): i\in V\mbox{ and } k\in\pa(i)\}$:
\begin{align} \label{ml-sem}
X_i=\bigvee\limits_{k \in \pa(i)} c_{ki} X_k \vee c_{ii} Z_i,\quad i=1, \ldots, d,
\end{align}
where $\pa(i)$ denotes the {parents} of node $i$ in $\D$ and $c_{ki}>0$ for $k\in \pa(i)\cup\{i\}$; the noise variables $Z_1,\ldots,Z_d$, represented by a generic random variable $Z$, are assumed to be { independent} and { identically distributed} with support $\R_+:=(0,\infty)$ and {\em regularly varying} with index $\al\in\R_+$, 
abbreviated by  $Z\in\text{RV}(\alpha)$. Denoting the distribution function of $Z$  by $F_Z$, the latter means  that 
\begin{align*}
\lim_{t\to\infty}\frac{1- F_Z(xt)}{1-F_Z(t)} = x^{-\al}
\end{align*}
for every $x\in\R_+$. Examples for $F_Z$ include Cauchy, Pareto, and log-gamma distributions. For details and background on regular variation, see e.g. \cite{Resnick1987,Resnick2007}.

The properties of the noise variables imply the existence of a normalizing sequence $a_n\in\R_+$ such that  for independent copies $\bfx^{(1)},\ldots, \bfx^{(n)}$ of $\bfx$, 
\begin{align}\label{maxdom}
{a}_n^{-1} \bigvee_{\nu=1}^n \bs{X}^{(\nu)} \std \bs{M}, \quad n\to\infty, 
\end{align}
where $\bs{M}$ is a non-degenerate random vector with distribution function denoted by $G$
and all operations are taken componentwise. Thus $\bfx$ is in the maximum domain of attraction of $G$; we write 
$\bfx\in\MDA(G)$. The limit vector $\bs{M}$ (its distribution function $G$)  is necessarily {max-stable}: for all $n\in\mathbb{N}$ and  independent copies $\bs{M}^{(1)},\ldots, \bs{M}^{(n)}$ of $\bs{M}$, the distributional equality $\boldsymbol{a}_n \bs{M}+\boldsymbol{b}_n\eqd \bigvee_{\nu=1}^n \bs{M}^{(\nu)}$  holds for appropriately chosen
 normalizing sequences $\boldsymbol{a}_n\in\R_+^d$ and $\boldsymbol{b}_n\in \R^d$. In  the present situation we have $\boldsymbol{a}_n=(n^{1/\al},\ldots,n^{1/\al})$ and $\boldsymbol{b}_n=(0,\ldots,0)$. Furthermore, $\bs{M}$ is again a \mSEM\ on $\D$, with the same weights in \eqref{ml-sem} as $\bfx$ and  standard $\al$-Fr{\'e}chet distributed noise variables, i.e., 
\begin{align*}
F_Z(x)=\Phi_\al (x)=\exp\{-x^{-\al}\}, \quad x\in\R_+.
\end{align*}
A proof of \eqref{maxdom} as well as an explicit formula for $G$ and its  univariate and bivariate marginal distributions can be found in Appendix~\ref{A2}, \autoref{prop2.1}. 

In what follows we summarize the most important properties of $\bfx$ presented in \cite{GK1} which are needed throughout the paper.  Every component of $\bfx$ can be written as a max-linear function of its ancestral noise variables: 
\begin{align} \label{ml-noise}
X_i=\bigvee\limits_{j \in \An(i)} b_{ji}Z_j,  \quad i=1,\ldots,d, 
\end{align}
where $\An(i)=\an(i)\cup\{i\}$ and $\an(i)$ are the { ancestors} of $i$ in $\D$ \cite[][Theorem~2.2]{GK1}. For $i\in V$,   $b_{ii}=c_{ii}$.  For $j\in\an(i)$,  $b_{ji}$ can be determined by a path analysis of $\D$ as explained in the following. Throughout we write $k\to i$  whenever $\D$ has an edge from $k$ to $i$. With every path  $p=[ j=k_0 \rightarrow k_1\rightarrow \ldots \rightarrow  k_n=i]$ we associate a weight, which we define to be the product of the edge weights along $p$ multiplied by $c_{jj}$. The coefficient $b_{ji}$ is then the maximum weight of all  paths from $j$ to $i$. In summary, we have for $i\in V$ and $j\in\an(i)$,
\begin{align} \label{bs}
b_{ji}=\bigvee\limits_{p \in P_{ji}} d_{ji}(p)\quad \text{with} \quad d_{ji}(p) :=  c_{k_0k_0}\prod_{\nu=0}^{n-1} c_{k_\nu k_{\nu+1}},
\end{align}
where $P_{ji}$ is the set of all paths from $j$ to $i$. 
For all $i\in V$ and $j\in V\setminus \An(i)$ we set $b_{ji}=0$. We call the coefficients  $b_{ji}$  {\em ML coefficients}  (\MLC s)  and summarize them in the {\em ML coefficient matrix} (\MLCM) $B = (b_{ij})_{d\times d}$.  
For the {\em reachability matrix} $R$ of $\D$, whose $ji$-th  entry is one if $j\in \An(i)$ and zero else, we find 
\begin{align}\label{reach}
R=\sgn(B),
\end{align}
where $\sgn$ denotes the signum function and is taken componentwise. As a consequence, the ancestors and descendants of every node in $\D$ can be obtained from $B$. 
 
Not all paths are needed for computing $b_{ji}$ in \eqref{bs}.  We call a path $p$ from $j$ to $i$ {\em max-weighted path from $j$ to $i$} if it realizes the maximum in \eqref{bs}, i.e., if $b_{ji}=d_{ji}(p)$. The concept of max-weighted paths is essential. This has been 
worked out in \cite{GK1}. For example, max-weighted paths may lead to more conditional independence relations  in the distribution of $\bfx$ than those encoded by $\D$ via the Markov property \citep[][Remark~3.9]{GK1}.  \mSEM s where all paths are max-weighted play a central role in this paper; we call them {\em  recursive max-weighted models} (\RMWM s).

Further \DAG s and weights may exist such that $\bfx$ satisfies \eqref{ml-sem}; for a detailed characterization of these \DAG s and weights, see~Theorem~5.4 of \citet[]{GK1}. The smallest \DAG\ 
 of this kind  is the one that has an edge $k\to i$ if and only if (iff) this is the only max-weighted path from $k$ to $i$ in $\D$ \cite[][Remark~5.2(ii) and Theorem~5.4(a)]{GK1}. We call this \DAG\ $\D^B$, the  {\em minimum \ML\ \DAG\ of $\bfx$}. It can be determined 
 from $B$ \cite[][Theorem~5.3]{GK1}.  The other \DAG s representing $\bfx$ in the sense of \eqref{ml-sem} are those that have at least the edges of $\D^B$ and the same reachability matrix. For edges contained in $\D^B$, the weights from \eqref{ml-sem}  are uniquely defined by $B$. From these weights the weights for the other edges can be derived. 

\brem\label{rem:spectral}
The random vector $\bfx$ and its distribution are characterized by the distribution $F_Z$ of the noise variables and the max-linear dependence structure 
induced by $\D$. So computing the max-stable limit distribution $G$ concerns only the marginal limits, whereas the max-linear dependence structure  remains always the same (cf. also the proof of \autoref{prop2.1}). This restrictive  dependence structure of $\bfx$ can be generalized naturally within the framework of {multivariate regular variation}.  See \cite{Resnick1987,Resnick2007} for background on multivariate regular variation.

In the literature various equivalent formulations of regular variation for random vectors can be found. The extent of a possible generalization can be probably best understood when considering an equivalent representation of the dependence in 
a regularly varying vector. A random vector $\bfx\in\R_+^d$ is 
regularly varying with index $\al\in\R_+$ iff  
there exists a random vector $\Theta$ with values in $\mathbb{S}^{d-1}=\{\bfsx\in\R_+^d : \|\bfsx\|=1\}$, where $\|\cdot\|$ is any norm in $\R_+^d$,
 such that for every $x\in\R_+$, 
\begin{align}\label{spectral}
\frac{\P(\|\bfx\|>tx, \bfx/\|\bfx\|\in \cdot)}{\P(\|\bfx\|>t)} \stv
x^{-\al} \P(\Theta\in\cdot), \quad t\to\infty. 
\end{align}
The notation $\stv$ stands for vague convergence on the Borel $\sigma$-algebra of $\mathbb{S}^{d-1}$. We immediately find from \eqref{spectral}  that the dependence structure of $\bfx$ is for moderate values of $\|\bfx\|$ arbitrary; only when  $\|\bfx\|$ becomes large, the dependence structure becomes that of $\Theta$. When assuming that  the dependence structure in the limit is max-linear given by $\D$ and the marginal limits are $\al$-Fr{\'e}chet (with an appropriate scale parameter), 
then $\bfx\in \MDA(G)$  
with $G$ still as in \autoref{prop2.1}; hence, $\bfx$ would have the same TDCs as in the present less general framework. So similarly to the flexibility of the margins, expressed by $Z\in\text{RV}(\al)$, there would  also be flexibility in the dependence structure.  

In this paper the restriction to the limiting max-linear dependence provides a sufficient model as the focus lies on the causal structure in terms of the \DAG s. 
This allows for a more concise notation and makes the focus of the paper 
more transparent.
\vspace{-\partopsep} 
\erem

\subsection*{The tail dependence matrix of $\bfx$}

For $i\in V$ we denote the distribution function of component $X_i$ of the \mSEM\ $\bfx$  
by $F_i$  and its generalized inverse by $F_i^{\leftarrow}(u)=\inf\{x\in\R_+: F(x)\ge u\}$ for $0<u<1$. 
The \TDC\ between $X_i$ and $X_j$ 
 is then given by the limit
\begin{align*}
\chi(i,j)=\lim_{u \uparrow 1} \P(X_i > F_i^{\leftarrow}(u)\mid X_j >F_j^{\leftarrow}(u)).
\end{align*}
We summarize all \TDC s 
 in the {\em tail dependence matrix} (\TDM) $\chi=(\chi(i,j))_{d\times d}$.

Defining the {\em standardized \MLCM}  of $\bfx$ by
\begin{align} \label{Bquer}
\B=\big(\b_{ij}\big)_{d\times d}:=\Big( \frac{b_{ij}^\al}{\sum_{k\in\An(j)} b_{kj}^\al}\Big)_{d\times d},
\end{align}
the \TDC\ between $X_i$ and $X_j$ can be computed as 
\begin{align}
\chi(i,j)=\chi(j,i)=\sum_{k\in\An(i) \cap \An(j)} \b_{k i} \wedge \b_{k j}.   \label{tdc}
\end{align}
By \eqref{reach} and \eqref{Bquer}  it is the sum of the pairwise minima of the $i$-th and $j$-th column of $\B$. A proof of \eqref{tdc} is given in Appendix~\ref{A2}.  There we  implicitly show that $\bfx$ and the limit vector $\bs{M}$ from \eqref{maxdom} have the same \TDM\ $\chi$. 

The \TDC\ $\chi(i,j)$ is zero iff $i$ and $j$ do not have common ancestors. Therefore, the {\em initial nodes} of $\D$ (i.e., the nodes  without parents
)  constitute a set $V_{\text{\smaller$0$}}$ of maximum cardinality such that  $\chi(i,j)$ is zero for all distinct $i,j\in V_{\text{\smaller$0$}}$. This property turns out to be helpful when identifying from $\chi$.
We also show that $\chi(i,j)$ is zero  iff  $X_i$ and $X_j$ are independent, which is reminiscent of the multivariate Gaussian distribution with its equivalence between  independence and zero correlation.   

Obviously, when investigating $\chi$, understanding the structure of $\B$ is essential. Not surprisingly, $\B$ inherits structural  properties from $B$.  For example, $\B$ is again a \MLCM\ of a \mSEM\ on the same \DAG\ $\D$, and its columns add up to one. Properties of $\B$, which we use throughout this paper, are summarized in Appendix~\ref{A1}, \autoref{propBquer}.

\subsection*{Identifiability from  $\chi$}

The main goal of this paper is to investigate how far the dependence structure of $\bfx$ and the \DAG\ $\D$ can be recovered from the \TDM\ $\chi$.   We call two \mSEM s that have the same \TDM\   {\em $\chi$-equivalent}. For example, $\bfx$ and the limit vector $\bs{M}$ from \eqref{maxdom} are $\chi$-equivalent. The set 
\begin{align}\label{Btilde}
\big\{  (\wt b_{ij})_{d\times d}\in \R_+^{d\times d}:\wt b_{ij}=\beta_j \b_{ij}^{1/\wt\alpha} \text{ for all $i,j\in V$  and $\beta_j\in \R_+$}\big\} 
\end{align}
contains the \MLCM s of all \mSEM s that have the same standardized \MLCM\ $\B$ as $\bfx$ and  regularly varying noise variables with index $\wt\alpha\in\R_+$; 
this can be verified  by using Theorem~5.7 of \cite{GK1}. Obviously, all the corresponding 
\mSEM s are also $\chi$-equivalent to $\bfx$. 
Therefore, given $\chi$ only, we can never identify the true representations \eqref{ml-sem} and \eqref{ml-noise} of $\bfx$ and the  
 \DAG\ $\D$. 

The \mSEM\ $\bfx$ has the same minimum \ML\ \DAG\ $\D^B$ as every \mSEM\ with \MLCM\ $\B$ (\autoref{propBquer}(e)). As a consequence,  $\D^B$ can be determined from $\B$ \citep[cf.][Theorem~5.3]{GK1}.  This raises the question of whether $\B$  and, hence, the minimum \ML\ \DAG\ of $\bfx$ are identifiable from $\chi$.  The answer is  generally no, quite simply due to the symmetry of $\chi$.

\bexam[$\B$ is not identifiable from $\chi$\label{exam:intro}]\\ Consider  two \mSEM s on the \DAG s $\D_1$ and $\D_2$ with  
 standardized \MLCM s 
\begin{multicols}{3}
\begin{minipage}{0.3\textwidth}
\hspace*{-1.25em}
\begin{tikzpicture}[->,every node/.style={circle,draw},line width=0.8pt, node distance=1.6cm,minimum size=0.8cm,outer sep=1mm]
\node (n1) {${1}$};
\node (n2) [right of=n1] {${2}$};
\node (n5)[draw=white,fill=white,left of=n1,node distance=0.88cm] {$\D_1$};
\foreach \from/\to in {n1/n2}
\draw[->, line width=0.8pt] (\from) -- (\to);
\end{tikzpicture}
\end{minipage}
\begin{minipage}{0.3\textwidth}
\vspace*{-1em}
\begin{align*}
\hspace*{-2.5em}
\B_1=
\begin{pmatrix}
1 & b\\
0& 1-b
\end{pmatrix}
\quad \text{and} \quad
\B_2=
\begin{pmatrix}
1-b & 0\\
b& 1
\end{pmatrix}
\end{align*}
\end{minipage}
\begin{minipage}{0.3\textwidth}
\hspace*{3em}
\begin{tikzpicture}[->,every node/.style={circle,draw},line width=0.8pt, node distance=1.6cm,minimum size=0.8cm,outer sep=1mm]
\node (n1) {${1}$};
\node (n2) [right of=n1] {${2}$};
\node (n5)[draw=white,fill=white,right of=n2,node distance=1cm] {$\D_2$};
\foreach \from/\to in {n2/n1}
\draw[->, line width=0.8pt] (\from) -- (\to);
\end{tikzpicture}
\end{minipage}
\end{multicols}
\noindent for some $b\in (0,1)$. 
For both we find the same \TDC s: $\chi(1,1)=\chi(2,2)=1$ and $\chi(1,2)=\chi(2,1)=b$. 
\eexam

We show, however, that $\B$ can be computed  recursively from $\chi$ and some additional information on the \DAG\ $\D$. This may be its reachability matrix $R$ but also only a {\em causal ordering} $\sigma$; i.e., $\sigma$ is a permutation on $V=\{1,\ldots,d\}$ such that $\sigma(j)<\sigma(i)$ for all $i\in V$ and $j\in\an(i)$. If $\bfx$ is max-weighted, then $\B$ is identifiable from $\chi$ and the initial nodes $V_{\text{\smaller$0$}}$ of $\D$.

The question also arises which \mSEM s are all $\chi$-equivalent to $\bfx$ 
and what their minimum \ML\ \DAG s are. Since   by \eqref{Btilde} every \MLCM\ of a \mSEM\ with \TDM\ $\chi$ can be obtained from its particular standardized version, it suffices to clarify which the standardized \MLCM s of all \mSEM s with \TDM\ $\chi$ are. To this end we use the identifiability results mentioned above to develop an algorithm that computes 
 these matrices  from $\chi$.  The proposed procedure can  be considerably simplified for \RMWM s. 

Another interesting point is how \DAG s of $\chi$-equivalent \mSEM s relate to each other.
Here we also investigate the \RMWM s as a relevant subclass of \mSEM s separately. 
For example, an initial node in a  \DAG\ of a \RMWM\  is again an initial node  in a \DAG\ of a $\chi$-equivalent \RMWM\ or it nust be a {terminal node} (i.e., a node without descendants). 

\medskip

Our paper is organized as follows. We provide some basic results in Section~\ref{s2}. For a \mSEM\ $\bfx$ 
we investigate its \TDM\  $\chi$ and link it to its standardized \MLCM\ $\B$ and its associated \DAG\ $\D$. 
Here we discuss the situations when two components of $\bfx$ have zero tail dependence. 
We also introduce the important concept of $\chi$-cliques, which allows us  to identify  potential initial node sets in $\D$ from $\chi$. 
Section~\ref{s4} is devoted to \RMWM s. We point out the specific properties of $\chi$ which lead to the identifiability of $\B$ from $\chi$ and the initial nodes. 
We also present necessary and sufficient conditions on a matrix to be the \TDM\ of a \RMWM. 
In Section~\ref{s3} we then 
study different identifiability problems 
based on $\chi$. We propose algorithms to compute $\B$ from $\chi$ and some further information on
$\D$ such as a causal ordering. 
We also explain how the standardized \MLCM s of all \mSEM s that have \TDM\ $\chi$ can be determined. In Section~\ref{s5} we consider $\chi$-equivalent \mSEM s and analyze relationships  between them and their \DAG s. We use these results to investigate whether \RMWM s on different \DAG s can be $\chi$-equivalent at all and if so under which conditions.  
Section~\ref{s6} concludes.

\medskip

Note that all recursion formulas presented in the paper are well-defined, since we work with \DAG s. Throughout we illustrate our findings with examples 
for  the (standardized) \MLCM\ of a \mSEM\ on a given \DAG. It can be verified by Theorem~4.2  or Corollary~4.3(a) of \cite{GK1}  that the presented matrices are indeed \MLCM s  of  \mSEM s on the particular \DAG s. Moreover, we use the following notation throughout the paper. 
We denote the  {ancestors, parents}, and {descendants} of  node 
  $i$ in $\D$ 
  by ${\an}(i)$, ${{\pa}}(i)$, and  ${\des}(i)$, respectively. We define $\An(i):={\an}(i)\cup\{i\}$,  ${\Pa}(i):={{\pa}}(i)\cup\{i\}$, and  ${\Des}(i):={\des}(i)\cup\{i\}$. For (possibly random) $a_i\in\R$  we set  $\bigvee_{i\in\emptyset}  a_i =0$ and
$\sum_{i\in\emptyset}  a_i =0$. We generally consider statements for $i\in\emptyset$ as invalid.

\section{The recursive max-linear model and its tail dependence matrix\label{s2}}

In this section for a \mSEM\ $\bfx$ on a \DAG\ $\D$,  we highlight some relations between its \TDM\ $\chi$, its standardized \MLCM\ $\B$, and the \DAG\  $\D$. 
They prove particularly useful when we identify the  \mSEM s that are $\chi$-equivalent to $\bfx$ in Section~\ref{s33}   or investigate \DAG s of $\chi$-equivalent \mSEM s in Section~\ref{s5}. 

\subsection{The tail dependence coefficients and max-weighted paths\label{s21}}

We start with lower and upper bounds for the \TDC\ between two components of $\bfx$ such that in $\D$ the two corresponding nodes  are connected by a path.  We also show that  max-weighted paths lead to simple expressions for the \TDC s and to nice relationships between them.  It is precisely these properties that motivate us to consider  \RMWM s in detail later on. 

\ble\label{propTC}
Let $i\in V$ and $j\in\an(i)$.
\begin{citemize}
\item[(a)] We have  
$ 0<\frac{\b_{ji}}{\b_{jj}} \le  \chi(j,i)$ with equality iff there is a max-weighted path from every $k\in\An(j)$ to $i$ passing through $j$. In that case, $\chi(i,j)= \sum_{k\in \An(j)} \b_{ki}$. 
\item[(b)]  We have 
 $\chi(i,j)\le \sum_{k\in \An(j)} \b_{k i}<1$. 
\item[(c)]  Let $k\in\des(j)\cap\an(i)$. If there is a max-weighted path from  every $\ell\in\An(j)$ to $k$ and from every  $\ell\in\An(j)$ to $i$ passing through $j$ as well as  from every $\ell\in\An(k)$ to $i$ passing through $k$, then
\begin{align}\label{propTC2}
 \chi(j,i)= \chi(j,k)\chi(k,i)<  \chi(j,k)\wedge \chi(k,i).
\end{align}
\end{citemize}
\ele

\bproof 
As $\An(j)\subseteq \An(i)$, we have by \eqref{tdc}, $\chi(j,i)=\sum_{k \in \An(j)} \b_{k i} \wedge \b_{k j}$.  \\ 
(a) For $k\in\An(j)$, by \autoref{propBquer}(d), (f),  $\frac{\b_{k j}\b_{ji}}{\b_{jj}} \le  \b_{k i} \wedge \b_{k j}$  with equality iff there is a max-weighted path from $k$ to $i$ passing through $j$.  With this, using also \autoref{propBquer}(b), (a), we obtain $\chi(j,i)\ge \frac{\b_{ji}}{\b_{jj}}\sum_{k\in\An(j)}\b_{k j}=\frac{\b_{ji}}{\b_{jj}}>0$ with equality iff there is a max-weighted path from every $k\in\An(j)$ to $i$ passing through $j$. 
 In that case  \autoref{propBquer}(d) yields $\chi(j,i)=  \sum_{k\in\An(j)} \frac{\b_{k j} \b_{ji}}{\b_{jj}}=\sum_{k\in\An(j)}\b_{k i}$.  \\
(b)  As $\An(j)\subsetneq \An(i)$, by \autoref{propBquer}(a), (b)  we find $\chi(j,i) \le \sum_{k \in \An(j)} \b_{k i} < \sum_{k \in \An(i)} \b_{k i}=1$. \\
(c)  The equality in \eqref{propTC2} follows from (a) and \autoref{propBquer}(d), the inequality then from the strict inequality in (b). 
\vspace{-\partopsep} 
\eproof

In the proof of \autoref{propTC} we have used that  for $i\in V$, $k\in\an(i)$, and $j\in\an(k)$, 
$\D$ has a max-weighted path from  $j$ to  $i$ passing through  $k$  iff  $\b_{ji}=\frac{\b_{jk}\b_{ki}}{\b_{kk}}$  (\autoref{propBquer}(d)).   
 As to the equality in \eqref{propTC2},  one could expect that the \MLC s can be replaced by the corresponding \TDC s. The following example disproves this. In particular, it proves that the converse of  \autoref{propTC}(c) is not true in general 
 and also that we may have the equality in \eqref{propTC2} although $k\not\in\des(j)\cap \an(i)$.  

\bexam[$\chi(j,i)=\chi(j,k)\chi(k,i)$ is neither necessary nor sufficient for $\b_{ji}=\frac{\b_{jk}\b_{ki}}{\b_{kk}}$\label{exam1}]
\begin{citemize}
\item[(1)] Consider a \mSEM\ on  $\D_1$ with standardized \MLCM\
\vspace*{-0.5em}
\begin{multicols}{2}
\begin{minipage}{0.945\textwidth}
\centering
\begin{align*}
\B=
\begin{pmatrix}
1 & 0&0.4&0.3\\
0 &1 &0.4&0.25\\
0&0&0.2&0.125\\
0&0&0&0.325\\
\end{pmatrix}.
\end{align*}
\end{minipage}
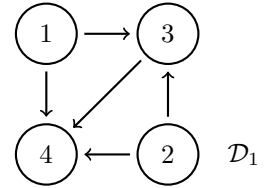
\begin{minipage}{0.5\textwidth}
\hspace*{10em}
\begin{tikzpicture}[->,every node/.style={circle,draw},line width=0.8pt, node distance=1.6cm,minimum size=0.8cm,outer sep=1mm]
\node (n1) {${1}$};
\node (n3) [below of=n1] {${4}$};
\node (n4)[right of=n1] {${3}$};
\node (n2) [below of=n4] {${2}$};
\node (n5)[draw=white,fill=white,right of=n2,node distance=1cm] {$\D_1$};
\foreach \from/\to in {n1/n3,n1/n4,n2/n3,n4/n3,n2/n4}
\draw[->, line width=0.8pt] (\from) -- (\to);
\end{tikzpicture}
\end{minipage}
\end{multicols}
\vspace*{-0.3em}
\noindent As $\b_{24}=\frac{\b_{23}\b_{34}}{\b_{33}}$,  the path $[2\to 3\to 4]$ is 
 max-weighted. 
 Computing $\chi$ we find $\chi(2,4)<\chi(2,3)\chi(3,4)$. That is, 
$\chi(2,4)\neq \chi(2,3)\chi(3,4)$ although there is a max-weighted path from $2$ to $4$ passing through $3$. 
\item[(2)] Now consider a \mSEM\ on $\D_1$ with standardized \MLCM\
\begin{align*}
\B=
\begin{pmatrix}
1 & 0&0.1&0.085\\
0 &1 &0.8&0.5\\
0&0&0.1&0.04\\
0&0&0&0.375\\
\end{pmatrix}.
\end{align*}
The path $[2 \to 3 \to 4]$ is not max-weighted, since $\frac{\b_{23}\b_{34}}{\b_{33}}\neq \b_{24}$. However, we have $\chi(2,3)\chi(3,4)=\chi(2,4)$. In summary,  $\chi(2,3)\chi(3,4)=\chi(2,4)$  although there is no max-weighted path from $2$ to $4$ passing through $3$. 
\item[(3)]  Finally, consider a  \mSEM\   on $\D_2$  
 with standardized \MLCM\ 
\vspace*{-0.5em}
\begin{multicols}{2}
\begin{minipage}{0.93\textwidth}
\centering
\begin{align*}
\B=
\begin{pmatrix}
1 & 0&1/3&1/6\\
0 &1 &1/3&1/3\\
0&0&1/3&0\\
0&0&0&1/2\\
\end{pmatrix}.
\end{align*}
\end{minipage}
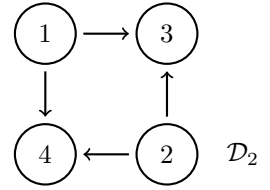
\begin{minipage}{0.5\textwidth}
\hspace*{10em}
\begin{tikzpicture}[->,every node/.style={circle,draw},line width=0.8pt, node distance=1.6cm,minimum size=0.8cm,outer sep=1mm]
\node (n1) {${1}$};
\node (n3) [below of=n1] {${4}$};
\node (n4)[right of=n1] {${3}$};
\node (n2) [below of=n4] {${2}$};
\node (n5)[draw=white,fill=white,right of=n2,node distance=1cm] {$\D_2$};
\foreach \from/\to in {n1/n3,n1/n4,n2/n3,n2/n4}
\draw[->, line width=0.8pt] (\from) -- (\to);
\end{tikzpicture}
\end{minipage}
\end{multicols}
\vspace*{-0.3em}
\noindent  Here we find $\chi(1,3)\chi(3,4)=\chi(1,4)$;
 but $3$ is not an ancestor of $4$.  According to this the equality in \eqref{propTC2} may hold although $k\not\in\des(j)\cap\an(i)$.
 \end{citemize}
\vspace{-5\partopsep} 
\eexam

\subsection{The tail dependence coefficients and the initial nodes}
  
In this section 
we mainly investigate how $\chi$ and  $\D$ relate to each other. 

Two components of $\bfx$ are independent iff the \TDC\ between them is zero.  We link  these two properties with the relationship between the two corresponding nodes in $\D$.

\bthe \label{chiequiv} 
Let $\bfx$ be a \mSEM\ on a \DAG\ $\D=(V,E)$   with \TDM\ $\chi$ and $i,j\in V$. 
Then the following statements are equivalent:
\begin{citemize}
\item[(a)]
$X_i$ and $X_j$ are independent.
\item[(b)]
$\An(i)\cap\An(j)=\emptyset$. 
\item[(c)]
$\chi(i,j)=0$. 
\end{citemize}
\ethe
\bproof
The equivalence between (a) and (b) follows  from representation \eqref{ml-noise} for $X_i$ and $X_j$ and the distributional
properties of the noise variables. The one between  (b) and (c) is immediate by  \eqref{tdc} and \autoref{propBquer}(a).  
\eproof

\brem\label{rem:cont_spectral} 
\begin{citemize}
\item[(i)]
 Let $R$ be the reachability matrix of $\D$. The $ij$-th ($ji$-th)  entry 
of $R^T R$ equals the cardinality of $\An(i)\cap \An(j)$. Thus by \autoref{chiequiv}, $\sgn(\chi)=\sgn(R^TR)$. That is, we learn from $\chi(i,j)>0$ only that $\An(i)\cap \An(j)\neq \emptyset$ but not whether $i$ and $j$ are connected by a path as is the case for the (standardized) \MLC s (\autoref{propBquer}(a) and  \eqref{reach}, respectively).
\item[(ii)] In the more general framework of \autoref{rem:spectral}, parts (a) and (b) of \autoref{chiequiv} would have to be replaced by 
\begin{citemize}
\item[(a')] $X_i$ and $X_j$ are asymptotically independent; i.e., the corresponding components of the limit vector in \eqref{maxdom} are independent. 
\item[(b')] The dependence structure in the limit is given by  a \DAG, in which $\An(i)\cap\An(j) = \emptyset$. 
\end{citemize}
The equivalence between (a') and (c) is a well-known result in extreme value theory; see e.g. Theorem~6.2.3 and the subsequent remark in \cite{DHF}.
\end{citemize}
\vspace{-5\partopsep} 
\erem

In what follows we investigate the relationship between $\chi$ and the initial nodes $V_{\text{\smaller$0$}}$ of $\D$. This is motivated by the fact that a \mSEM\ is recursively  defined  by the structure of $\D$. For example, to obtain representation \eqref{ml-noise} of $\bfx$ from its  representation \eqref{ml-sem} recursively, we would start with representation \eqref{ml-noise} of the components $X_i$ with $i\in V_{\text{\smaller$0$}}$. Then by proceeding iteratively we would substitute  the parental 
variables in \eqref{ml-sem}  by their representation \eqref{ml-noise}. Such an iterative procedure starting with the initial nodes could also identify all  \mSEM s which have (the given) \TDM\ $\chi$. 

The \TDC\ between two components of $\bfx$ simplifies considerably when in $\D$ one of the corresponding nodes is an initial node. If both nodes are initial nodes, then the \TDC\ between them is  zero. We provide these and further related results.  

\ble \label{lemW}
\begin{citemize}
\item[(a)] For distinct $i,j\in V_{\text{\smaller$0$}}$,  $\chi(i,j)=0$.
\item[(b)] Let $W\subseteq V$ such that $\chi(i,j)=0$ for all distinct  $i,j \in W$. Then $\vert W\vert \leq \vert V_{\text{\smaller$0$}}\vert$.
\item[(c)] For $i\in V$ and  $j\in V_{\text{\smaller$0$}}$,   $\An(i)\cap V_{\text{\smaller$0$}}=\{k\in V_{\text{\smaller$0$}}: \chi(k,i)>0\}$ and $\Des(j)=\{k \in V: \chi(j,k)>0\}$.  
\item[(d)] For $i\in V$ and $j\in V_{\text{\smaller$0$}}$, $\chi(j,i)=\b_{ji}$. 
\end{citemize}
\vspace{-\partopsep} 
\ele
\bproof
(a) and (c) follow from the fact that initial nodes have no ancestors and \autoref{chiequiv}. \\
(b) Assume that  $\vert W\vert > \vert V_{\text{\smaller$0$}} \vert$. Since for every $i\in V$ there is some $j\in \An(i)\cap V_{\text{\smaller$0$}}$,  we have $j\in\An(i_1)\cap\An(i_2)$ for  some $j \in V_{\text{\smaller$0$}}$ and distinct $i_1,i_2\in W$. As $\An(i_1) \cap \An(i_2) \neq \emptyset$, again by  \autoref{chiequiv}, $\chi(i_1,i_2)\neq 0$. This is, however, a contradiction to the fact that $\chi(i_1,i_2)=0$ as  $i_1,i_2\in W$. 
Hence, $\vert W\vert \le \vert V_{\text{\smaller$0$}} \vert$.\\
(d)  As $\An(j)=\{j\}$, we obtain from \eqref{tdc} by  \autoref{propBquer}(a), (f), 
$\chi(j,i)=\sum_{k=1}^d\b_{ki}\wedge\b_{kj} =\b_{ji}\wedge \b_{jj}=\b_{ji}$.
\eproof

From \autoref{lemW}(a), (b) we learn that $V_{\text{\smaller$0$}}$ is one of the node sets of maximum cardinality such that for every two distinct nodes,   the \TDC\ between their corresponding components of $\bfx$ is zero. We introduce a concept which allows us to determine  these sets from $\chi$ by a graph. For an illustration of these notions, we refer to \autoref{exam:AlgChinc3} below. 

\bde \label{chiclique} 
Let $\chi$ be the \TDM\ of a \mSEM\  on a \DAG\ $\D$. 
\begin{citemize}
\item[(a)] We call the undirected graph that has nodes $V$ and an edge between $k$ and $i$ iff $\chi(k,i)>0$, {\em $\chi$-graph}. 
\end{citemize}
Let  $\D^{\chi}$ be the complement of the $\chi$-graph  \citep[for the definition of the complement of an (undirected) graph, see e.g.][Chapter~1.1]{Diestel:2010} and $W\subseteq V$. 
\begin{citemize}
\item[(b)] We call $W$  a {\em $\chi$-clique} if it is a clique  in $\D^\chi$  \citep[for the definition of a clique in a 
graph, see e.g.][Definition~2.13]{KF}. 
\item[(c)]  We call  $W$    a {\em maximum $\chi$-clique} if it is a maximum clique of $\D^\chi$; i.e., $W$ is a clique in $\D^\chi$ such that no clique in $\D^\chi$ with higher 
cardinality exists.
\end{citemize}
\vspace{-5\partopsep} 
\ede

The $\chi$-graph associated with the \TDM\ $\chi$ of $\bfx$ corresponds to the {covariance graph} of the random vector $\bfx$ introduced in \citet{cox1993}, in which two (distinct) nodes are  connected by an edge iff their corresponding components are dependent (cf. \autoref{chiequiv}). In the non-Gaussian case, however, the name covariance graph is misleading.

The following theorem is an immediate consequence of Definition~\ref{chiclique} and \autoref{lemW}(a), (b). 

\bthe\label{RemMaxCl}  
Let $\bfx$ be a \mSEM\ on a \DAG\ $\D$ with \TDM\ $\chi$. Then the set $V_{\text{\smaller$0$}}$ is a maximum $\chi$-clique.
\ethe

\autoref{RemMaxCl} raises the question of how $V_{\text{\smaller$0$}}$ is related to possible other maximum $\chi$-cliques.

\ble \label{le:Ws}
Let $W$ be a  maximum $\chi$-clique. 
\begin{citemize} 
\item[(a)]  There is only one bijection $\varphi:V_{\text{\smaller$0$}}\to W$ such that for every $j\in V_{\text{\smaller$0$}}$,  $\chi(j,\varphi(j))>0$  and $\chi(j,i)=0$ for all $i\in W\setminus \{\varphi(j)\}$.  
\item[(b)] 
Let $\varphi$ be the bijection from (a). Then for  $j\in V_{\text{\smaller$0$}}$,  $\An(\varphi(j))\cap V_{\text{\smaller$0$}}=\{j\}$ and $\Des(j)\cap W=\{\varphi(j)\}$. In particular,  if $j\neq \varphi(j)$, then $\D$ has a path from $j$ to $\varphi(j)$. 
\item[(c)] Let $i,j\in V\setminus W$. If $\chi(i,j)< \sum_{k\in W} \chi(k,i)\wedge \chi(k,j)$, then $V_{\text{\smaller$0$}}\neq W$. 
\end{citemize}
\vspace{-\partopsep} 
\ele
\bproof
(a) Since maximum $\chi$-cliques have the same cardinality, we know from \autoref{RemMaxCl} that $\vert V_{\text{\smaller$0$}} \vert =\vert W\vert$.  As for every $i\in W$, $\An(i)\cap  V_{\text{\smaller$0$}}\neq \emptyset$, it suffices by \autoref{lemW}(c) to show that  $\vert \Des(j)\cap W\vert=1$ for  $j\in V_{\text{\smaller$0$}}$. We first assume that $\vert\Des(j)\cap W \vert>1$. Using \autoref{chiequiv} similarly as in the proof of \autoref{lemW}(b) yields  a contradiction. Hence, $\vert\Des(j)\cap W \vert\le 1$.  As $\vert V_{\text{\smaller$0$}}\vert =\vert W\vert$,  $\vert \Des(j)\cap W\vert=1$ must hold.\\
(b) follows from (a) and \autoref{lemW}(c). \\
(c)  Assume that $V_{\text{\smaller$0$}}=W$. Using \autoref{propBquer}(a) and \autoref{lemW}(d)  we obtain from \eqref{tdc}
\begin{align*}
\chi(i,j)=\sum_{k=1}^d \b_{k i}\wedge \b_{k j}\ge\sum_{k\in W} \chi(k,i)\wedge\chi(k,j).
\end{align*} 
Since this contradicts the conditions of (c), $V_{\text{\smaller$0$}}$ and $W$ must be different. 
\vspace{-\partopsep} 
\eproof

\section{The recursive max-weighted model and its tail dependence matrix\label{s4}}

In this section we focus on \RMWM s, i.e.,  
 \mSEM s where all paths are max-weighted. We first present some structural properties of a \RMWM\ $\bfx$ on a \DAG\ $\D$ with standardized \MLCM\ $\B$. We then investigate its \TDM\ $\chi$ and show that 
 the assumption of all paths in $\D$ being max-weighted  involves simple relations  between the \TDC s and the (standardized) \MLC s.  Finally, we give necessary and sufficient conditions on a matrix to be the \TDM\ of a \RMWM\ on a given \DAG.

\subsection{Some structural properties of a recursive max-weighted model\label{s41}}

All \mSEM s on polytrees are \RMWM s simply because in a polytree there is at most one path between every two (distinct) nodes  \citep[see~also][Example~3.2]{GK1}. Furthermore, a \RMWM\ can be constructed on every \DAG, as the following example shows. Note the particularly simple structure of the \mSEM\ introduced by it.

 \bexam\label{hom}[The homogeneous model] \\ 
Let $\D=(V,E)$ be a \DAG\ with $V=\{1,\ldots,d\}$ and $Z_1,\ldots,Z_d$  as in \eqref{ml-sem}. 
Consider the \mSEM\ defined by 
\begin{align*}
 X_i := \frac1{\vert \An(i)\vert^{1/\alpha}}{\Big(\bigvee\limits_{k \in \pa(i)} \vert \An(k)\vert^{1/\alpha} X_{k} \vee Z_i\Big)},\quad i=1,\ldots,d.
\end{align*} 
We find that every path $p$ from $j$ to $i$ has the same weight $d_{ji}(p)= \vert \An(i)\vert^{-1/\alpha}$. As a consequence, every path is max-weighted and $\bfx$ is a \RMWM.  Its representation \eqref{ml-noise} is given by 
\begin{align*}
 X_i &=\frac{1}{ \vert \An(i)\vert^{1/\alpha} } \bigvee\limits_{j\in \An(i)} Z_j, \quad i=1,\ldots,d.
\end{align*}
\noindent 
 For the \TDC\  from \eqref{tdc} between $X_i$ and $X_j$, we have
\begin{align*}
\chi(i,j)= \sum_{k\in\An(i)\cap\An(j)} \frac{1}{\vert \An(i)\vert}  \wedge \frac{1}{\vert \An(j)\vert} =\frac{\vert \An(i)\cap \An(j)\vert}{\vert \An(i)\vert\vee\vert\An(j)\vert }.
\end{align*}
If $j\in\an(i)$, then this reduces to $\chi(i,j)=\vert \An(j)\vert/\vert\An(i)\vert$. Finally, by \autoref{prop2.1}  the components  of the limit vector $\bs{M}$ introduced in \eqref{maxdom} are standard $\al$-Fr{\'e}chet distributed. 
 \eexam

Recall from the Introduction the prominent role of the minimum \ML\ \DAG\ $\D^B$ of $\bfx$, which equals the minimum \ML\ \DAG\  $\D^\B$ of a \mSEM\ with \MLCM\ $\B$ (\autoref{propBquer}(e)). The fact that $\bfx$ is max-weighted ensures that $\D^\B$ only depends on $\sgn(\B)$  but not on the precise values of the  standardized \MLC s. Since $\sgn(\B)$ is the reachability matrix of $\D$ (\eqref{reach} and \autoref{propBquer}(a)), $\D^\B$  can be determined from pure graph theoretical properties. 
 To clarify this we introduce a basic concept in graph theory, which goes back to \citet{Aho1972}.

\bde\label{defi2.7}
Let $\D$ be a \DAG. 
 \begin{citemize}
\item[(a)] An edge $k\to i$ is {redundant} if $\D$ has another path from $k$ to $i$. 
\item[(b)] The \DAG\ $\D^{\tr}$ obtained from $\D$ by deleting its redundant  edges  is called { transitive reduction} of $\D$.
\end{citemize}
\vspace{-5\partopsep} 
\ede

Since $\D^\B$ has an edge $k\to i$  iff this is the only max-weighted path from $k$ to $i$ in $\D$, the fact that  $\D$ has only max-weighted paths yields part (i) of the following remark.   By \autoref{defi2.7} and \autoref{propBquer}(a)  (ii) is a consequence of (i). 
 
\brem\label{transred} 
Let $\D^\tr$ be the transitive reduction of $\D$.
\begin{citemize}
\item[(i)] The \DAG s $\D^\B$ and $\D^{\tr}$ coincide. 
\item[(ii)] $\D^\B$ is the \DAG\ with the minimum number of edges that has reachability matrix $\sgn(\B)$.  
\item[(iii)] Even if  $\bfx$  is a \mSEM\ but not max-weighted, 
 it may happen that $\D^\B=\D^{\tr}$ with all paths max-weighted in $\D^\B$. In that case all results presented in this section hold with respect to $\D^\tr$. 
\end{citemize}
\vspace{-5\partopsep} 
\erem

\subsection{Properties of the tail dependence coefficients of a recursive max-weighted model\label{s42}}

The following result  points out  the simple structure of $\chi$. It follows from \autoref{propTC}(a), (c), since in $\D$ all paths are max-weighted. 

\ble\label{chimwm} Let $i\in V$. 
\begin{citemize}
\item[(a)] For $j \in \An(i)$, $\chi(j,i)=\frac{\b_{ji}}{\b_{jj}}=\sum_{k\in\An(j)}\b_{ki}=\sum_{k\in\An(j)}\b_{kk}\chi(k,i)$.
\item[(b)] For $k\in\an(i)$ and $j\in\an(k)$, $\chi(j,i)=\chi(j,k)\chi(k,i)< \chi(j,k)\wedge\chi(k,i)$.
\item[(c)] For $j\in\an(i)$ and some path $[j=k_0 \to k_1\to \cdots \to k_n=i]$, $\chi(j,i)=\prod_{\nu=0}^{n-1}\chi(k_\nu,k_{\nu+1})$.
\end{citemize}
\ele
 
The equality $\chi(j,i)=\chi(j,k)\chi(k,i)$ for some $j\in\An(i)\cap\An(k)$ does not necessarily imply that  $k\in\An(i)$ (cf. part (3) of \autoref{exam1}). For \RMWM s, however,  whenever these products hold for all $j\in\An(i)\cap\An(k)\cap V_{\text{\smaller$0$}}$, where  $V_{\text{\smaller$0$}}$ are again the initial nodes in $\D$, we can conclude that $k\in\An(i)$. 
 
\bpr\label{chimwm1}
For $i,k\in V$, $k\in \An(i)$ iff $\chi(j,i)=\chi(j,k)\chi(k,i)$ for all $j\in \An(i)\cap\An(k)\cap V_{\text{\smaller$0$}}$. 
\vspace{-\partopsep} 
\epr
\bproof
Assume that $\chi(j,i)= \chi(j,k) \chi(k,i)$ for all $j\in \An(i)\cap \An(k)\cap V_0$. We first show that $\chi(\ell,i)\le \chi(\ell,k)$ for every 
$\ell\in\An(i)\cap\An(k)$. 
We obtain for $j\in\An(\ell)\cap V_{\text{\smaller$0$}}$, using the assumptions and \autoref{chimwm}(b),  
\begin{align*}
\chi(k,i)=\frac{\chi(j,i)}{\chi(j,k)}=\frac{\chi(j,\ell)\chi(\ell,i)}{\chi(j,\ell)\chi(\ell,k)}=\frac{\chi(\ell,i)}{\chi(\ell,k)}. 
\end{align*}
Hence,  $\chi(\ell,i)=\chi(\ell,k)\chi(k,i)$ and $\chi(\ell,i)\le \chi(\ell,k)$. 
Together with \autoref{chimwm}(a) we then find from \eqref{tdc}
\begin{align*}
\chi(k,i)=\sum_{\ell\in\An(k)\cap\An(i)}\b_{\ell \ell} (\chi(\ell, k) \wedge\chi(\ell, i)) =\sum_{\ell\in\An(k)\cap\An(i)}\b_{\ell \ell} \chi(\ell, i)=\chi(k,i) \sum_{\ell\in\An(k)\cap\An(i)}\b_{\ell \ell} \chi(\ell, k).
\end{align*}
By the assumptions and \autoref{chiequiv}  $\chi(k,i)>0$ so that $\sum_{\ell\in\An(k)\cap\An(i)}\b_{\ell \ell} \chi(\ell, k)=1$. 
As $1=\sum_{\ell\in\An(k)}\b_{\ell\ell}\chi(\ell,k)$  (\autoref{chimwm}(a))  and $\b_{\ell\ell}\chi(\ell,k)>0$ for all $\ell\in\An(k)$ (\autoref{propBquer}(a) and \autoref{chiequiv}), we have $\An(i)\cap\An(k)=\An(k)$. This finally implies that  $\An(k)\subseteq \An(i)$, equivalently $k\in\An(i)$.

The converse statement holds due to \autoref{chimwm}(b). 
\eproof

In \autoref{chimwm}(a) we have written  the  positive standardized \MLC s  as  functions of themselves and \TDC s.   We now present expressions for  them only in terms of \TDC s. 

\bpr \label{BwithChi} 
For $i\in V$ and  $j\in\An(i)$,
\begin{align}\label{BwithChi:eq}
\b_{ji}= \chi(j,i)- \sum_{k\in \an(j)}\lambda_{jk} \chi(k,i) \quad \text{with\quad $\lambda_{jk}=1-\sum_{\ell \in \des(k) \cap \an(j)} \lambda_{j\ell}$}.
\end{align} 
\epr
\bproof
As by \autoref{chimwm}(a)  $\b_{ji}= \chi(j,i)-\sum_{k\in\an(j)} \b_{ki}$, it suffices to show that $\sum_{k\in \an(j)}\lambda_{jk} \chi(k,i)=\sum_{k\in\an(j)} \b_{ki}$. Using again \autoref{chimwm}(a) yields
\begin{align*}
\sum_{k\in \an(j)}\lambda_{jk}\chi(k,i)=\sum_{k\in \an(j)} \lambda_{jk}\sum_{\ell\in\An(k)} \b_{\ell i}.
\end{align*}
Noting that $k\in \an(j)$ and $\ell\in\An(k)$ iff  $\ell\in \an(j)$ and $k\in\Des(\ell)\cap\an(j)$, we can interchange the two summation operators  to obtain
\begin{align*}
\sum_{k\in \an(j)} \lambda_{jk}\sum_{\ell\in\An(k)} \b_{\ell i} &=\sum_{\ell\in \an(j)} \b_{\ell i} \sum_{k\in\Des(\ell)\cap\an(j)} \lambda_{jk}=\sum_{\ell\in \an(j)} \b_{\ell i} \big(\lambda_{j\ell}+\sum_{k\in\des(\ell)\cap\an(j)} \lambda_{jk}\big)=\sum_{\ell\in \an(j)} \b_{\ell i},
\end{align*}
where we have used the definition of $\lambda_{j\ell}$ for the last equality. 
\eproof

Before we give an example of representation \eqref{BwithChi:eq}, we summarize some characteristics of the coefficients $\lambda_{jk}$. 
Denoting by $\pa^\tr(j)$ the parents of $j$ in the transitive reduction $\D^\tr$ of $\D$,   we  have $\lambda_{jk}=1$ for $k\in\pa^\tr(j)$ as $\des(k)\cap\an(j)=\emptyset$. For  $k\in\an(j)\setminus \pa^\tr(j)$ it can be verified that $\lambda_{jk}\neq 0$ iff there exists no $\wt k\in\des(k)\cap\an(j)$ such that  $\vert \Des(\wt k)\cap \pa^\tr(j)\vert = \vert \Des(k)\cap \pa^\tr(j)\vert $.

\bexam[On representation \eqref{BwithChi:eq}\label{exam:lambda}]\\ 
Consider a \RMWM\ $\bfx$ on the \DAG\ $\D$ depicted below, and note that here $\D=\D^\tr$. We determine, as an example, representation   \eqref{BwithChi:eq} for   the  \MLC s $\b_{36,66}$ and $\b_{98,99}$: 
\begin{align*}
\b_{36,66}=\chi(36,66)-\chi(35,66), \quad \b_{98,99}=\chi(98,99)- \chi(34,99)-\chi(66,99)-\chi(97,99)+\chi(2,99)+\chi(35,99).
\end{align*} 
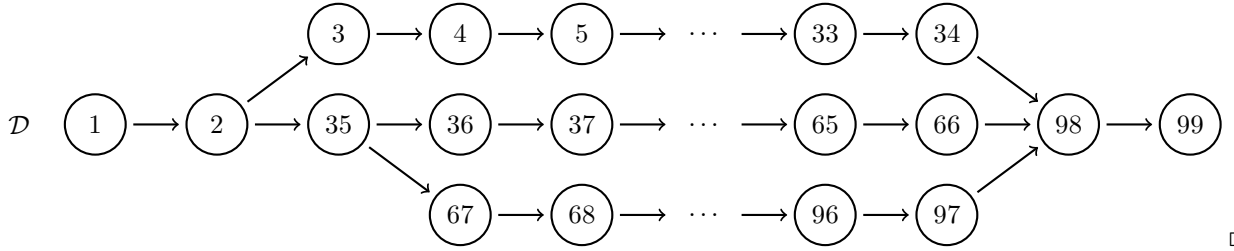
\begin{center}
\hspace*{-1.5em}
\begin{tikzpicture}[->,every node/.style={circle,draw},line width=0.8pt, node distance=1.6cm,minimum size=0.8cm,outer sep=1mm]
\node (1)  {$1$};
\node (2) [right of=1]  {$2$};  
\node (35) [right of=2]  {$35$};  
\node (36) [right of=35]  {$36$};  
\node (37) [right of =36]{$37$};
\node (38) [draw=white,right of =37]{$\ldots$};
\node (65) [right of =38]{$65$};
\node (66) [right of =65]{$66$};
  \node (98) [right of =66]{$98$};
  \node (99) [right of =98]{$99$};  
  \node (3) [above of=35,node distance=1.2cm]  {$3$};  
    \node (4) [above of=36,node distance=1.2cm]  {$4$};      
        \node (5) [above of=37,node distance=1.2cm]  {$5$};      
    \node (6) [draw=white,above of =38,node distance=1.2cm]{$\ldots$};
\node (33) [above of =65,node distance=1.2cm]{$33$};  
\node (34) [above of =66,node distance=1.2cm]{$34$};  
\node (67) [below of =36,node distance=1.2cm]{$67$};  
\node (68) [right of =67]{$68$};
\node (69) [draw=white,right of =68]{$\ldots$};
\node (96) [right of =69]{$96$};  
\node (97) [below of =66,node distance=1.2cm]{$97$};  
\node (n4)[draw=white,fill=white,left of=1,node distance=1.0cm] {$\D$};
    \foreach \from/\to in {1/2,2/35,35/36,36/37,37/38,38/65,65/66,66/98,98/99,2/3,3/4,4/5,5/6,6/33,33/34,34/98,35/67,67/68,68/69,69/96,96/97,97/98}
  \draw (\from) -- (\to);
\end{tikzpicture}
\end{center}
\vspace{-10\partopsep} 
\eexam

We address again the interrelations between the \TDC s and prove that every \TDC\  can  be written as  linear combination of  minima of two \TDC s. 

\bpr\label{chistr}
For $i,j\in V$,
\begin{align}
\chi(i,j)&=\sum_{k \in \An(i) \cap \An(j)} \mu_{ij,k} (\chi(k,i) \wedge \chi(k,j)) \quad \text{with\quad $\mu_{ij,k}=1-\sum_{\ell \in \des(k) \cap \An(i) \cap \An(j)} \mu_{ij,\ell}$}.\label{lambdadef}
\end{align}
\epr
\bproof  
Applying \autoref{chimwm}(a) and \autoref{propBquer}(b), (d) we obtain for $k\in\An(i)\cap\An(j)$, 
\begin{align*}
\chi(k,i) \wedge \chi(k,j)&= \frac{\b_{ki}}{\b_{kk}}\wedge\frac{\b_{kj}}{\b_{kk}}=   \Big(\frac{\b_{ki}}{\b_{kk}}\wedge\frac{\b_{kj}}{\b_{kk}}\Big)\Big(\sum_{\ell \in \An(k)} \b_{\ell k} \Big) =\sum_{\ell \in \An(k)} \frac{\b_{\ell k}\b_{ki}}{\b_{kk}}  \wedge \frac{\b_{\ell k}\b_{kj}}{\b_{kk}}  =\sum_{\ell \in \An(k)} \b_{\ell i} \wedge \b_{\ell j}.
\end{align*}
With this we then have
\begin{align*} 
\sum_{k \in \An(i) \cap \An(j)} \mu_{ij,k} (\chi(k,i) \wedge \chi(k,j)) 
&=\sum_{k \in\An(i)\cap\An(j)} \mu_{ij,k}  \sum_{\ell \in \An(k)} \b_{\ell i} \wedge \b_{\ell j}. 
\end{align*}
Using that  $k\in\An(i)\cap\An(j)$ and $\ell\in\An(k)$ iff $\ell\in\An(i)\cap\An(j)$ and $k\in \Des(\ell)\cap\An(i)\cap\An(j)$ to interchange the summation operators  similarly  as  in the proof of \autoref{BwithChi} and the definition of $\mu_{ij,\ell}$ similarly as the one of  $\lambda_{j\ell}$ there, we finally find \eqref{lambdadef}.  
\eproof

For $i,j\in V$ denote by $\text{lca}(i,j)$ the lowest common ancestors of $i$ and $j$; i.e., $k\in\text{lca}(i,j)$ iff $k\in\An(i)\cap\An(j)$ and $\D$ has no path from $k$ to another node in $\An(i)\cap\An(j)$. For $\mu_{ij,k}$ from \eqref{lambdadef} we have $\mu_{ij,k}=1$ 
for $k\in\text{lca}(i,j)$ as in that case 
 $\des(k)\cap\An(i)\cap\An(j)=\emptyset$. 
 It can be verified  that 
$\mu_{ij,k}= 0$  for  $k\in (\An(i)\cap\An(j))\setminus \text{lca}(i,j)$ iff there exists some $\wt k\in\des(k)\cap\An(i)\cap\An(j)$ such that  $\vert \Des(\wt k)\cap \text{lca}(i,j)\vert = \vert \Des(k)\cap  \text{lca}(i,j)\vert $.
With this,   if $j\in\An(i)$, then $\mu_{ij,j}=1$ and $\mu_{ij,k}=0$ for $k\in\an(j)$. Thus in that case the right-hand side of the first equality in \eqref{lambdadef} is equal to $\chi(j,i)\wedge \chi(j,j)=\chi(j,i)$, and representation \eqref{lambdadef} is  trivial.  Note the analogy of the coefficients $\mu_{ij,k}$   to the coefficients $\lambda_{jk}$ in \eqref{BwithChi:eq}.

\bexam[On representation \eqref{lambdadef}]\\
Consider a \RMWM\ on the \DAG\ $\D$ depicted below. We present, as an example, representation \eqref{lambdadef} for the  \TDC s $\chi(95,96)$ and $\chi(96,97)$: 
\begin{align*}
\chi(95,96)=&\chi(33, 95)\wedge \chi(33,96 ),\\
\chi(96,97)=& \chi(33,96)\wedge \chi(33,97)+ \chi(64,96)\wedge \chi(64,97)+\chi(94,96)\wedge \chi(94,97)\\ 
& -\chi(34,96)\wedge \chi(34,97)-\chi(2,96)\wedge \chi(2,97). 
\end{align*}
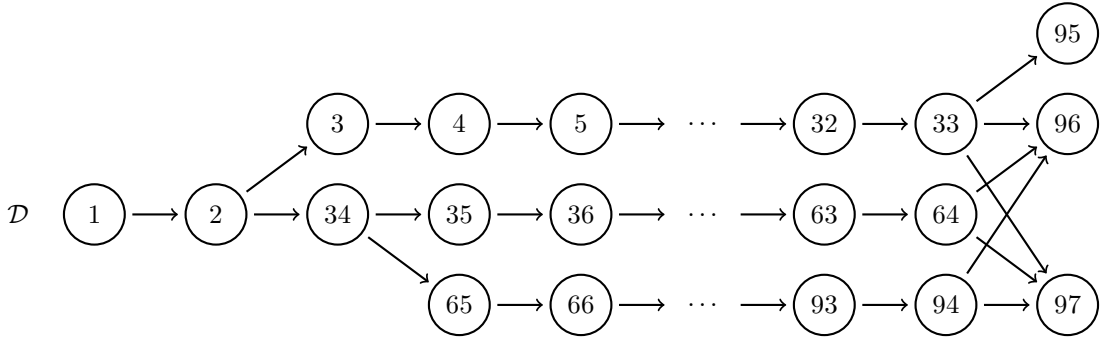
\begin{center}
\vspace*{-2.75em}
\hspace*{-2em}
\begin{tikzpicture}[->,every node/.style={circle,draw},line width=0.8pt, node distance=1.6cm,minimum size=0.8cm,outer sep=1mm]
\node (1)  {$1$};
\node (2) [right of=1]  {$2$};  
\node (35) [right of=2]  {$34$};  
\node (36) [right of=35]  {$35$};  
\node (37) [right of =36]{$36$};
\node (38) [draw=white,right of =37]{$\ldots$};
\node (65) [right of =38]{$63$};
\node (66) [right of =65]{$64$};
  \node (98) [right of =34]{$96$};
  \node (99) [right of =97]{$97$};  
  \node (3) [above of=35,node distance=1.2cm]  {$3$};  
    \node (4) [above of=36,node distance=1.2cm]  {$4$};      
        \node (5) [above of=37,node distance=1.2cm]  {$5$};      
    \node (6) [draw=white,above of =38,node distance=1.2cm]{$\ldots$};
\node (33) [above of =65,node distance=1.2cm]{$32$};  
\node (34) [above of =66,node distance=1.2cm]{$33$};  
\node (67) [below of =36,node distance=1.2cm]{$65$};  
\node (68) [right of =67]{$66$};
\node (69) [draw=white,right of =68]{$\ldots$};
\node (96) [right of =69]{$93$};  
\node (97) [below of =66,node distance=1.2cm]{$94$};
\node (100) [above of =98,node distance=1.2cm]{$95$};    
\node (n4)[draw=white,fill=white,left of=1,node distance=1.0cm] {$\D$};
    \foreach \from/\to in {34/100,1/2,2/35,35/36,36/37,37/38,38/65,65/66,66/98,2/3,3/4,4/5,5/6,6/33,33/34,34/98,35/67,67/68,68/69,69/96,96/97,97/98,34/99,66/99,97/99} 
  \draw (\from) -- (\to);
\end{tikzpicture}
\end{center} 
\vspace{-10\partopsep} 
\eexam

We conclude this section with necessary and sufficient conditions on a matrix to be  the \TDM\ of a \RMWM\ on a given \DAG\ $\D$.  To be such a matrix,
the $ij$-th ($ji$-th)  entry of the matrix must satisfy a property depending on the relationship between  $i$ and $j$ in $\D$. For example, based on \autoref{chiequiv}, it must be zero iff $\An(i)\cap\An(j)=\emptyset$. By \autoref{propBquer}(e), \autoref{transred}(i), and Theorem~5.4 of \cite{GK1}, a \RMWM\ on $\D$ is a \RMWM\  on every \DAG\ that has reachability matrix $R$ of $\D$. Consequently, it would be sufficient to specify  $R$ and  to require the four conditions below for any \DAG\ with reachability matrix $R$ such as  the transitive reduction $\D^\tr$ of  $\D$. 

\bthe \label{th:chartdm}  
Let $\D=(V,E)$ be a \DAG\ with nodes $V=\{1,\ldots,d\}$ and reachability matrix $R$. Let $\chi=(\chi(i,j))_{d\times d}$ be a symmetric matrix with ones on the diagonal. For $i\in V$ define  
 $\b_{ii}:=1-\sum_{k\in\an(i)}\b_{kk}\chi(k,i)$ recursively. Then $\chi$ is the \TDM\ of a  \RMWM\ $\bfx$ on $\D$ iff the following conditions hold:
\begin{citemize}
\item[(a)] $\sgn(\chi)=\sgn(R^TR)$.
\item[(b)] For all $i\in V$, $\b_{ii}>0$.
\item[(c)] For all $i\in V$, $j \in\an(i)$, and $k\in\des(j)\cap\pa(i)$, $\chi(j,i)=\chi(j,k)\chi(k,i)$.
\item[(d)] For all $i,j\in V$ such that   $i\not\in \An(j)$ and $j\not\in\An(i)$ but $\An(i)\cap\An(j)\neq\emptyset$,
\end{citemize}
\begin{align*}
\chi(i,j)= \sum_{k\in\An(i)\cap\An(j)}  \b_{kk} (\chi(k,i)\wedge\chi(k,j) ).
\end{align*} 
In that case $\b_{ii}$ is the $i$-th diagonal entry of the standardized \MLCM\  $\B$ of $\bfx$. Furthermore,  for $i,j\in V$,  $\b_{ji}=0$ if $j\in V\setminus\An(i)$, and $\b_{ji}=\b_{jj}\chi(j,i)$ if $j\in\an(i)$. 
\vspace{-\partopsep} 
\ethe
\bproof 
Assume that $\chi$ is the \TDM\ of a \RMWM\ $\bfx$ on $\D$.   The statements (a) and (c) follow  from \autoref{rem:cont_spectral}(i) and  \autoref{chimwm}(b). 
By \autoref{chimwm}(a)  $\b_{ii}$ is the $i$-th diagonal entry of the standardized \MLCM\ $\B$ of $\bfx$. Since all $\b_{ii}$ are positive according to \autoref{propBquer}(a),  assertion
 (b) holds.  
The representation of $\chi(i,j)$ in (d) is again a consequence of 
\autoref{chimwm}(a). 

Assume now that (a)-(d) hold.
 For  every $i\in V$ define  $\b_{ji}:=\b_{jj}\chi(j,i)$ for all $j\in\an(i)$ and  for  all $j\in V\setminus\An(i)$, $\b_{ji}:=0$. We first show that $\B=(\b_{ij})_{d\times d}$ is  the  \MLCM\ of a \RMWM\ on $\D$, where weights from its representation \eqref{ml-sem} are given by  $c_{ii}:=\b_{ii}$  and $c_{ki}:=\frac{\b_{ki}}{\b_{kk}}=\chi(k,i)$  for $i\in V$ and $k\in\pa(i)$.  As $\sgn(\chi)=\sgn(R^TR)$  and $\b_{ii}>0$, the weights $c_{ki}$ for $i\in V$ and $k\in\Pa(i)$ are positive,  which is a necessary condition for them by the definition of a \mSEM\ in \eqref{ml-sem}. 
Let $p=[j=k_0\to k_1\to\ldots\to k_n=i]$ be a path in $\D$. Using (c) iteratively yields
\begin{align*}
d_{ji}(p)&=c_{jj}\prod_{\nu=0}^{n-1}c_{k_\nu,k_{\nu+1}}=\b_{jj}\prod_{\nu=0}^{n-1} \chi(k_\nu,k_{\nu+1})
=  \b_{jj} \chi(j,k_2)\prod_{\nu=2}^{n-1} \chi(k_\nu,k_{\nu+1})=\ldots 
=\b_{jj}\chi(i,j)=\b_{ji}. 
\end{align*}
This implies that  $\B=(\b_{ij})_{d\times d}$  is the \MLCM\ of a \RMWM\ $\bfx$. 
Since it suffices to specify one \mSEM\ that has \TDM\ $\chi$, we may assume that $Z\in\text{RV}(1)$. 
Denoting the \TDM\ of $\bfx$ by $\ov\chi=(\ov\chi(i,j))_{d\times d}$, it remains to show that $\ov\chi=\chi$.  Since the diagonal entries of $\chi$ equal one, the equality of the diagonal entries is obvious.  For  $i,j\in V$ such that $\An(i)\cap\An(j)=\emptyset$, the $ij$-th ($ji$-th) entries of $\chi$ and $\ov\chi$ are zero and, hence,  equal due to condition (a) and  \autoref{chiequiv}.  The matrix $\B$ is the standardized \MLCM\ of $\bfx$  as  $\al=1$ and $\b_{ii}=1-\sum_{k\in\an(i)}\b_{ki}$ for every $i\in V$.  Thus for $i\in V$ and  $j\in\an(i)$ we have by \autoref{chimwm}(a) and the  definition of $\B$ that  $\ov\chi(j,i)=\frac{\b_{ji}}{\b_{jj}}=\chi(i,j)$. Finally, for  $i,j\in V$ such that $j\not\in\An(i)$ and $i\not\in \An(j)$ but $\An(i)\cap\An(j)\neq\emptyset$, using  \autoref{chimwm}(a), the result shown before,  and condition (d), we obtain 
\begin{align*}
\ov\chi(i,j)&= \sum_{k \in \An(i) \cap \An(j)} \b_{kk}  (\ov\chi(k,i)\wedge\ov\chi(k,j))  =\sum_{k\in\An(i)\cap\An(j)}  \b_{kk}  (\chi(k,i)\wedge\chi(k,j)) =\chi(i,j).
\vspace{-\partopsep}
\end{align*}
\eproof

In \autoref{exam:AlgChinc1} below we present a possible application of \autoref{th:chartdm}.

\brem 
In \autoref{th:chartdm} the coefficients $\b_{ii}$  can also be defined by $1-\sum_{k\in\an(i)}\lambda_{ik} \chi(k,i)$ with $\lambda_{ik}$ as in \eqref{BwithChi:eq}.  We give a sketch of a proof of this  assertion: we  show that $\lambda_{ik}=1-\sum_{\ell\in\des(k)\cap\an(i)}\lambda_{\ell k}$ and use this to verify 
that  if (c) holds, then the assertion is valid as well. 
 Moreover, condition (d) can  be replaced by 
\begin{citemize}
\setlength{\itemindent}{1em}
\item[(d')] For all $i,j\in V$ such that   $i\not\in \An(j)$ and $j\not\in\An(i)$ but $\An(i)\cap\An(j)\neq\emptyset$,
\end{citemize}
\begin{align*}
\chi(i,j)=\sum_{k \in \An(i) \cap \An(j)} \mu_{ij,k} (\chi(k,i) \wedge \chi(k,j)) \quad  \text{with\quad $\mu_{ij,k}$ as in \eqref{lambdadef}}.
\end{align*}
By going through the proof of \autoref{th:chartdm}, we observe that this can be done due to the representation of $\chi(i,j)$ in \eqref{lambdadef}. 
\vspace{-\partopsep} 
\erem

  \section{Identifiability problems  based on the tail dependence matrix of a recursive max-linear model\label{s3}} 

Throughout this section we assume that the \TDM\ $\chi$ of a \mSEM\ $\bfx$ on a \DAG\ $\D$ with standardized \MLCM\ $\B$ is given. 
 We first show the identifiability of $\B$ from $\chi$ and the reachability matrix $R$  of $\D$. We then assume that the reachability relation of $\D$ is not fully known but only a causal ordering $\sigma$. 
This still leads to identifiability of $\B$ from $\chi$. 
We also investigate whether  $\B$  can be recovered from $\chi$ and the initial nodes $V_{\text{\smaller$0$}}$ of $\D$. 
 It turns out that this is generally not possible, 
 but we verify it for \RMWM s. We prove the different identifiability results by providing   algorithms which compute $\B$ from $\chi$ and the additionally known information on $\D$.  Finally, based on these results we present an approach, which finds the standardized \MLCM s of all \mSEM s with \TDM\ $\chi$. 
 Since this method simplifies  for \RMWM s considerably, we give an adapted and modified version for this subclass of \mSEM s.

\subsection{Identifiability from the tail dependence matrix and the reachability matrix\label{s30}}

The following algorithm computes $\B$  from  $\chi$  and  $R$  recursively.  The rows of $\B$ are filled up successively until $\B$ is obtained, where 
the number of ancestors determines the order  in which the rows are treated. The existence of such an algorithm  proves the identifiability of  $\B$ 
 from $\chi$ and $R$.

\balg[Find $\B$ from $\chi$ and $R$\label{BChialg1}]\\
For $\nu=0,\ldots,d-1$,\\
\hspace*{2em} for $j\in V$ such that $\vert\an(j)\vert=\nu$, set
\begin{align}\label{eq:BquerFromChi}
\b_{ji}=0 \,\,\, \text{for all $i\in V\setminus\Des(j)$}\quad   \text{and}\quad \b_{ji}= \chi(j,i)-\sum_{k\in\an(j)} \b_{ki}\wedge\b_{kj} \,\,\, \text{for all $i\in\Des(j)$}.
\end{align}
\ealg

Eq.~\eqref{eq:BquerFromChi} follows from \autoref{propBquer}(a), \eqref{tdc}, and \autoref{propBquer}(f). If $\bfx$ is max-weighted, then by \autoref{chimwm}(a) 
\eqref{eq:BquerFromChi} can be replaced by 
\begin{align}\label{eq:BChi1}
\b_{ji}=0\,\,\, \text{for all $i\in V\setminus\Des(j)$}, \quad \b_{jj}
=1-\sum_{k\in\an(j)} \b_{kj},\quad \text{and}\quad \b_{ji}=\b_{jj}\chi(j,i)\,\,\, \text{for all $i\in\des(j)$}.
\end{align}
To avoid the iterative loop, we can also use \eqref{BwithChi:eq} for computing the diagonal entries of $\B$. Note, however, that this requires to calculate the coefficients $\lambda_{jk}$ appearing in \eqref{BwithChi:eq}  recursively as well. 

\subsection{Identifiability from the tail dependence matrix and a causal ordering\label{s31}}

So far we have dealt with the identifiability from $\chi$ and the reachability matrix $R$ of $\D$. Here we investigate the identifiability from $\chi$ and a causal ordering $\sigma $ of $\D$. If $R$ is given, then we know for every two (distinct) $i,j \in V$ whether there is a path from $j$ to $i$; but from $\sigma$ we only learn that there is no path from $j$ to $i$ if $\sigma(j)>\sigma(i)$. 
 
There exists a causal ordering for every \DAG\ due to the acyclicity   \citep[see also][Appendix~A]{Diestel:2010}. However, it is not necessarily unique. For example, the \DAG\ $\D_1$ from \autoref{exam1} has the identity function on $V=\{1,2,3,4\}$ and  the permutation $\wt\sigma$ on $V$ given by $\wt\sigma(2)=1,\,\wt\sigma(1)=2,\, \wt\sigma(3)=3,\, \wt\sigma(4)=4$ 
as causal orderings. 

The \DAG\ $\D$ has a causal ordering which can be completely described by its initial nodes $V_{\text{\smaller$0$}}$ and  $\chi$ as follows.

\ble\label{RemMaxCl1} 
We denote the initial nodes by $V_{\text{\smaller$0$}}=\{i_1,\ldots,i_{\vert V_{\text{\smaller$0$}}\vert}\}$ and define $V_{\text{\smaller$0$}}^i:=\{k\in V_{\text{\smaller$0$}}:\chi(k,i)>0\}$ for $i\in V$.  Then $\D$ has a causal ordering $\sigma$  such that 
\begin{align} \label{red:co}
\sigma(i_\nu)=\nu\,\,\, \text{for $\nu=1,\ldots,\vert V_{\text{\smaller$0$}}\vert$} \quad \text{and} \quad  \text{for all $i,j\in V$,}\,\,\, \sigma(j)<\sigma(i)\,\,\,\text{whenever $\vert V_{\text{\smaller$0$}}^j\vert < \vert V_{\text{\smaller$0$}}^i\vert$}. 
\end{align}
\ele
\bproof
Recall from \autoref{lemW}(c) that $V_{\text{\smaller$0$}}^j=\An(j)\cap V_{\text{\smaller$0$}}$ and  $V_{\text{\smaller$0$}}^i=\An(i)\cap V_{\text{\smaller$0$}}$. With this it is not difficult to see that $\D$ has such a causal ordering. 
\eproof

Now we give an iterative procedure which computes  $\B$ from $\chi$ and   $\sigma$. Obviously, this proves the identifiability of $\B$ from $\chi$ and $\sigma$. Here the rows of $\B$ are also filled up successively, where the order of the nodes given by $\sigma$ defines the order in which the rows are treated.  
 
\balg[Find $\B$ from $\chi$ and $\sigma$\label{BChiAlg}]\\
For $\nu=1,\ldots,d$,\\ 
\hspace*{2em} for $j\in V$ such that $\sigma(j)=\nu$, set
\begin{align}
   \begin{aligned}
 \b_{ji}&=0 \,\,\, \text{for all $i\in V$ such that $\sigma(j)> \sigma(i)$,}  \label{topB:eq}\\
  \b_{ji}&= \chi(j,i)-\sum\limits_{k:\, \sigma(k)<\sigma(j)}\b_{ki}\wedge \b_{kj} \,\,\, \text{for all $i\in V$ such that $\sigma(j)\le\sigma(i)$.}\\
   \end{aligned}
\end{align}
\ealg

Eq.~\eqref{topB:eq} can be obtained from \eqref{tdc} by using \autoref{propBquer}(a), the definition of a causal ordering, and \autoref{propBquer}(f).

\subsection{Identifiability of recursive max-weighted models from the tail dependence matrix and the initial nodes\label{s32}}

In what follows we assume $\bfx$  to be max-weighted. Then recalling  \autoref{lemW}(c), \autoref{chimwm1} involves a procedure to determine $R$   from $\chi$ and $V_{\text{\smaller$0$}}$. Since \autoref{BChialg1} computes $\B$  from $\chi$ and $R$,  we can identify  $\B$ from $\chi$ and $V_{\text{\smaller$0$}}$. This is usually not possible outside the class of \RMWM s.

\bexam[$\B$ is generally not identifiable from $\chi$ and $V_{\text{\smaller$0$}}$\label{examV0ng}]\\
Consider two \mSEM s on $\D_1$ and $\D_2$ with standardized \MLCM s 
$\B_{1}$ and $\B_2$ given by
\vspace*{-0.65em}
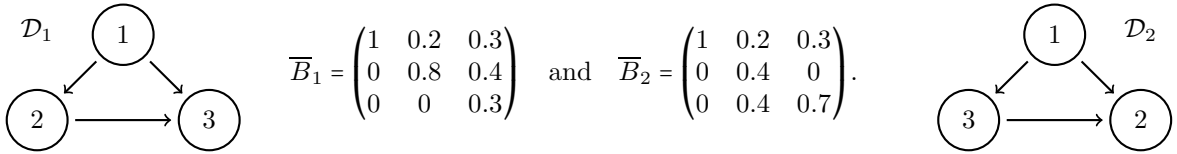
\begin{multicols}{3}
\begin{minipage}{0.4\textwidth}
\flushleft
\begin{tikzpicture}[->,every node/.style={circle,draw},line width=0.8pt, node distance=1.6cm,minimum size=0.8cm,outer sep=1mm]
\node (n1) {${1}$};
\node (n2) [below left of=n1] {${2}$};
\node (n3)[below right of=n1] {${3}$};
\node (n4)[draw=white,fill=white,above of=n2,node distance=1.2cm] {$\D_1$};
\foreach \from/\to in {n1/n2,n1/n3,n2/n3}
\draw[->, line width=0.8pt] (\from) -- (\to);
\end{tikzpicture}
\end{minipage}
\begin{minipage}{0.2\textwidth}
\centering
\begin{align*}
\hspace*{-4.2em}
\B_{1} =
\begin{pmatrix}
1 & 0.2&0.3\\
0 &0.8 &0.4\\
0&0&0.3
\end{pmatrix}
\quad \text{and} \quad
\B_{2}=
\begin{pmatrix}
1 & 0.2&0.3\\
0 &0.4 &0\\
0&0.4&0.7
\end{pmatrix}.
\end{align*}
\end{minipage}
\hspace*{-0.5em}
\begin{minipage}{0.27\textwidth}
\flushright
\begin{tikzpicture}[->,every node/.style={circle,draw},line width=0.8pt, node distance=1.6cm,minimum size=0.8cm,outer sep=1mm]
\node (n1) {${1}$};
\node (n2) [below left of=n1] {${3}$};
\node (n3)[below right of=n1] {${2}$};
\node (n5)[draw=white,fill=white,above of=n3,node distance=1.2cm] {$\D_2$};
\foreach \from/\to in {n1/n2,n1/n3,n2/n3}
\draw[->, line width=0.8pt] (\from) -- (\to);
\end{tikzpicture}
\end{minipage}
\end{multicols}
\vspace*{-0.3em}
\noindent   
We find by \autoref{propBquer}(d)  
that none of the two models is max-weighted. 
Since both have the same $\chi$ and $\D_1$ and $\D_2$ share the same initial node 
$V_{\text{\smaller$0$}}=\{1\}$, we cannot distinguish between $\B_1$ and $\B_2$ based on $\chi$ and $V_{\text{\smaller$0$}}$.  
\eexam

Proceeding as suggested by  \autoref{chimwm1} to recover $R$ from $\chi$ and $V_{\text{\smaller$0$}}$ is very tedious,  since many conditions may need to be verified. 
Therefore, we introduce an alternative method  
which computes $\B$ from $\chi$ and $V_{\text{\smaller$0$}}$: we first determine a causal ordering $\sigma$ of  $\D$ and apply then \autoref{BChiAlg} to obtain $\B$.  From the next proposition we learn how  a causal ordering $\sigma$ of $\D$ can be computed from $\chi$ and $V_{\text{\smaller$0$}}$; note that we encountered property (i) in \eqref{red:co}.

\bpr\label{IdentifyMwm} 
Let  $V_{\text{\smaller$0$}}^i$ for $i\in V$  be as in \autoref{RemMaxCl1}. 
Every permutation  $\sigma$ on $V$ such that for all $i,j\in V$, 
\begin{citemize}
\item[(i)] $\sigma(j)<\sigma(i)$ whenever $\vert V_{\text{\smaller$0$}}^j\vert < \vert V_{\text{\smaller$0$}}^i\vert$ and
\item[(ii)] $\sigma(j)<\sigma(i)$ whenever  $\vert V_{\text{\smaller$0$}}^j\vert=\vert V_{\text{\smaller$0$}}^i\vert$ and $\max_{k\in V_{\text{\smaller$0$}}^i}\chi(k,i)<\max_{k\in V_{\text{\smaller$0$}}^j}\chi(k,j)$
\end{citemize}
is a causal ordering of $\D$.  
\vspace{-\partopsep} 
\epr

\bproof 
Assume that $\sigma$ is no causal ordering of $\D$, i.e.,  $\sigma(j)>\sigma(i)$ for some $i\in V$ and $j\in \an(i)$.  
Recall from \autoref{lemW}(c) that  $V_{\text{\smaller$0$}}^j=\An(j)\cap V_{\text{\smaller$0$}}$ and $V_{\text{\smaller$0$}}^i=\An(i)\cap V_{\text{\smaller$0$}}$. As $j\in\an(i)$,   $V_{\text{\smaller$0$}}^j \subseteq  V_{\text{\smaller$0$}}^i $. But then because of the properties of $\sigma$,  $V_{\text{\smaller$0$}}^j= V_{\text{\smaller$0$}}^i$  and $\max_{k \in V_{\text{\smaller$0$}}^j}\chi(k ,j)\le \max_{k \in V_{\text{\smaller$0$}}^j}\chi(k ,i)$. 
Assume now that $j\in V^j_{\text{\smaller$0$}}$, and note that  $i\not\in V^j_{\text{\smaller$0$}}$ as $j\in\an(i)$. 
Then, since for $i_1,i_2\in V$ the \TDC\ $\chi(i_1,i_2)=1$ iff $i_1=i_2$ (cf.~\eqref{tdc} and \autoref{propBquer}(a)), 
we find $1=\max_{k \in V_{\text{\smaller$0$}}^j}\chi(k ,j)\le \max_{k \in V_{\text{\smaller$0$}}^j}\chi(k ,i) <1$.   
This contradiction proves that   $j\not\in V^j_{\text{\smaller$0$}}$, which implies again that $V_{\text{\smaller$0$}}^j=\an(j)\cap V_{\text{\smaller$0$}}$. As   $\max_{k \in V_{\text{\smaller$0$}}^j}\chi(k ,j)\le \max_{k \in V_{\text{\smaller$0$}}^j}\chi(k ,i)$, 
 $\chi(k,j)\le \chi(k,i)$  for some $k \in  \an(j)\cap V_{\text{\smaller$0$}}$. Observe from \autoref{chimwm}(b) that $j\not\in\an(i)$, since otherwise $\chi(k,i)< \chi(k,j)$. This, however, contradicts our original assumption, and $\sigma$ must be a causal ordering of $\D$. 
\eproof

Finally, we clarify the precise steps of our approach to determine $\B$ from $\chi$ and $V_{\text{\smaller$0$}}$.

\balg[Modification of \autoref{BChiAlg} for \RMWM s: find $\B$ from $\chi$ and  $V_{\text{\smaller$0$}}$\label{BChiV0}]
\begin{citemize}
\item[1.] Find a causal ordering $\sigma$ of $\D$ from $\chi$ and $V_{\text{\smaller$0$}}$:\\
for $\nu=1,\ldots,\vert V_{\text{\smaller$0$}}\vert$,
 \begin{citemize}
\item[] find all $j\in V$ such that  $\vert V^j_{\text{\smaller$0$}}\vert=\vert\{k\in V_{\text{\smaller$0$}}:\chi(k,j)>0\}\vert=\nu$ and summarize them in the set $A_\nu$;
\item[] sort the nodes $k_1,\ldots,k_{\vert A_\nu\vert}$ from $A_\nu$ such that 
\begin{align*}
\max_{\ell\in V_{\text{\smaller$0$}}}\chi(\ell,k_1)\, \ge \, \max_{\ell\in V_{\text{\smaller$0$}}}\chi(\ell,k_2) \, \ge \,  \ldots \, \ge\, \max_{\ell\in V_{\text{\smaller$0$}}}\chi(\ell,k_{\vert A_\nu\vert});
\end{align*}
for $\mu=1,\ldots, \vert A_\nu\vert$,
\begin{citemize}
\item[] set $\sigma(k_\mu)=\sum_{\lambda=1}^{\nu-1}\vert A_\lambda\vert +\mu$, where  $\sum_{\lambda=1}^{0}:=0$.
\end{citemize}
\end{citemize}
\item[2.] Apply \autoref{BChiAlg} to obtain  $\B$ from $\chi$ and $\sigma$.  
\end{citemize}
\ealg

Observe from \autoref{IdentifyMwm} that every permutation $\sigma$ on $V$ which can be chosen in step 1. is indeed a causal ordering of $\D$. 

\subsection{Identifiability from the tail dependence matrix\label{s33}}

We now combine the previous results to find the standardized \MLCM s of  all \mSEM s that have \TDM\ $\chi$. 
In the first part we deal with general \mSEM s. Because of the identifiability properties derived in Section~\ref{s32},    we assume in the second part that 
$\chi$ is the \TDM\ of a \RMWM.   We provide an algorithm, which outputs the standardized \MLCM s of all \RMWM s that have \TDM\ $\chi$. 

\subsubsection*{(General) recursive max-linear models}

Every permutation $\wt\sigma$ on $V=\{1,\ldots,d\}$ is a causal ordering of a \DAG\ with nodes $V$ but not necessarily of a \DAG\ that corresponds to a \mSEM\ with \TDM\ $\chi$. But if this is the case, then  applying \autoref{BChiAlg} with $\sigma=\wt\sigma$ yields the corresponding standardized \MLCM\ $\B$. This suggests the following procedure to prove 
the existence of a \mSEM\ which has \TDM\ $\chi$ and whose associated \DAG\ has causal ordering $\wt\sigma$: first apply  \autoref{BChiAlg} with $\sigma=\wt\sigma$,  and check then whether the obtained matrix $\B$ is the standardized  \MLCM\ of a \mSEM\ which has \TDM\ $\chi$ and whose associated \DAG\ has causal ordering $\wt\sigma$. In the second step it is enough to verify that $\B$  is the \MLCM\ of a \mSEM, which can be done by Theorem~5.7 of \cite{GK1}. 

\ble \label{Bstand+chi} 
Let $\wt\sigma$ be a permutation on $V$ and  $\B$ the matrix obtained by applying \autoref{BChiAlg} with $\sigma=\wt\sigma$.  If  $\B$ is the \MLCM\ of a \mSEM\ (\RMWM), then $\B$ is  the  standardized \MLCM\ of a \mSEM\ (\RMWM) which has \TDM\ $\chi$ and whose associated  \DAG\ has causal ordering $\wt\sigma$.  
\vspace{-\partopsep} 
\ele
\bproof
Let $\bfx$ be the \mSEM\ (\RMWM) with \MLCM\ $\B$ and $Z\in\text{RV}(1)$. Its existence is guaranteed as $\B$ is the \MLCM\ of a \mSEM\ (\RMWM).  We show that $\bfx$ has standardized \MLCM\ $\B$ and \TDM\ $\chi$ as well as that its associated \DAG\ $\D$ has causal ordering $\wt\sigma$.  Recall from \eqref{reach} that   $\sgn(\B)$ is the reachability matrix of $\D$. Thus by  \eqref{topB:eq} $\wt\sigma$ is a causal ordering of $\D$ and  $\b_{ii}=1-\sum_{k\in\an(i)}\b_{ki}$ for every $i\in V$. As the latter holds and  $\al=1$,  $\B$ is  the standardized \MLCM\ of $\bfx$. The fact that $\bfx$ has \TDM\ $\chi$ also follows from \eqref{topB:eq}.  
\eproof

Lemma~\ref{Bstand+chi}  suggests  a ``naive" method to find the standardized \MLCM s of all  \mSEM s that have \TDM\ $\chi$: 
 for every permutation on $V$ compute the matrix $\B$ from \autoref{BChiAlg}, and check whether  it is the \MLCM\ of a \mSEM; if so, then $\B$ is the standardized \MLCM\ of a \mSEM\ with \TDM\ $\chi$.  
 However, the number of permutations on $V$ to be investigated can often 
be significantly reduced.  By \autoref{RemMaxCl} and \autoref{le:Ws}(c) the set of all maximum $\chi$-cliques $W$ (see~\autoref{chiclique}) such that $\chi(i,j)\ge \sum_{k\in W} \chi(k,i)\wedge \chi(k,j)$ for all $i,j\in V\setminus W$ contains the initial node sets 
of all \DAG s underlying \mSEM s with \TDM\ $\chi$. 
So it suffices to investigate the causal orderings of \DAG s that have such initial nodes $W$.  
 But also the number of causal orderings to be investigated for every such set $W$ can be reduced further  by  \autoref{RemMaxCl1}: it is enough to consider those permutations on $V$, which satisfy the properties $\sigma$ has in  \eqref{red:co}  with $V_{\text{\smaller$0$}}=W$. 
The following algorithm  describes the precise steps of an approach to find the standardized \MLCM s of all \mSEM s with \TDM\ $\chi$.

\balg[Find all $\B$ from $\chi$\label{alg:Bconsistent}]
\begin{citemize}
\item[1.] Find  all maximum $\chi$-cliques: 
\begin{citemize}
\item[(a)] find the complement $\D^\chi$ of the $\chi$-graph; 
\item[(b)] find all maximum cliques of $\D^\chi$.
\end{citemize}
\item[2.] 
For every maximum $\chi$-clique $W=\{i_1,\ldots,i_{\vert W\vert}\}$,
\begin{citemize}
\item[(a)] check   $\chi(i,j)\ge \sum_{k\in W} \chi(k,i)\wedge \chi(k,j)$ for all $i,j\in V\setminus W$;\\ 
 if not, then there is  no  \mSEM\ with \TDM\ $\chi$ on a \DAG\ with initial nodes $W$; \\
else,
\begin{citemize}
\item[(b)] for every permutation $\wt\sigma$ on  $V=\{1,\ldots,d\}$ such that 
\begin{citemize}
\setlength{\itemindent}{1em}
 \item[] $\wt\sigma(i_\nu)=\nu$ for $\nu=1,\ldots,\vert W\vert$ and
 \item[] $\wt\sigma(j)<\wt\sigma(i)$ whenever $\vert\{k\in W:\chi(k,j)>0\}\vert <  \vert\{k\in W:\chi(k,i)>0\}\vert$,
  \end{citemize}
\begin{citemize}
\item[i.] 
apply \autoref{BChiAlg} with $\sigma=\wt\sigma$; 
\item[ii.] check whether $\B$ obtained in i. 
 is the \MLCM\ of a \mSEM; for instance using  Theorem~5.7 of \cite{GK1}; \\
if not, then there is no \mSEM\ with  \TDM\ $\chi$ on a \DAG\ with causal ordering~$\wt\sigma$;\\
 else,  $\B$ is the  standardized \MLCM\ of a  \mSEM\ with \TDM\ $\chi$. 
  \end{citemize}
\end{citemize}
\end{citemize}
\end{citemize}
\ealg 

When the algorithm returns a standardized \MLCM\ $\B$ of a \mSEM\ with \TDM\ $\chi$ in step~ii., then it is not necessary  to perform steps i., ii. for further permutations on $V$ which are causal orderings of  \DAG s with reachability matrix $\sgn(\B)$, since all of them would lead to the same $\B$. For the application of  \autoref{alg:Bconsistent}, we have assumed so far that $\chi$ is the \TDM\ of a \mSEM. If this is not the case,  \autoref{alg:Bconsistent} would not produce any output. The same applies to \autoref{alg:BconsistentMW} below if $\chi$ is not the 
 \TDM\ of a \RMWM. 

 One could drop  step~2.(a) and perform step~2.(b) for all maximum $\chi$-cliques. 
 However, the performance of step~2.(a) can be very effective.

\bexam[Not all maximum $\chi$-cliques are initial node sets\label{exam:alg:red}]\\
Consider the \TDM\ $\chi$ of a \mSEM\ on the \DAG\ $\D$ depicted below.  Note that such a \mSEM\ is max-weighted, since $\D$ is a polytree (cf.~Section~\ref{s41}). 
\autoref{chiequiv} yields that  the sets $\{1\}, \ldots,\{1000\}$ are the maximum $\chi$-cliques. 
For  every $k\in \{2,\ldots,999\}$  we know from  \autoref{chimwm}(b)  that  $\chi(1,1000)
< \chi(1,k)\wedge \chi(k,1000)$. The property tested in step~2.(a) is therefore not fulfilled for the maximum $\chi$-cliques $W\in \{\{2\},\ldots,\{999\}\}$. 
However,  we can  verify by \autoref{chimwm}(b) that it is fullfilled for $W\in \{\{1\},\{1000\}\}$.  
Consequently,  step~2.(b) needs only be performed for $W\in \{\{1\}, \{1000\}\}$ and not for the other $998$ maximum $\chi$-cliques.
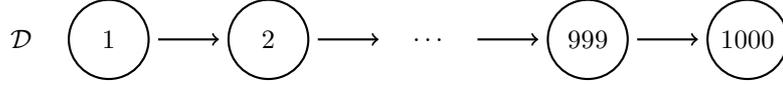
\vspace*{-0.3em}
\begin{center}
\begin{tikzpicture}[->,every node/.style={circle,draw},line width=0.8pt, node distance=2.1cm,minimum size=1.05cm,outer sep=1.25mm]
   \node (1)  {$1$};
      \node (2) [right of=1]  {$2$};  
        \node (3) [draw=white,right of =2]{$\ldots$};
         \node (4) [right of =3]{$999$};
               \node (5) [right of =4]{$1000$};
               \node (n5)[draw=white,fill=white,left of=1,node distance=1.15cm] {$\D$};
    \foreach \from/\to in {1/2,2/3,3/4,4/5}
  \draw (\from) -- (\to);
\end{tikzpicture}
\end{center}
\vspace{-10\partopsep} 
\eexam

It is indeed necessary to perform step~ii., i.e.,  to verify that a matrix $\B$ obtained in i. is a \MLCM\ of a \mSEM.

\bexam[Not every $\B$ obtained in ii. belongs to a \mSEM]\\
Consider the \TDM
\begin{align*}
\chi=\begin{pmatrix}
1	&	1/10	 & 1/3\\
1/10	& 1	& 13/30\\
1/3 & 13/30 & 1
\end{pmatrix}.
\end{align*}
\noindent Performing steps~i.~and~ii. of \autoref{alg:Bconsistent} with  $\wt\sigma$ being the identity function on $V=\{1,2,3\}$ and also with $\wt\sigma$ given by  $\wt\sigma(1)=1,\, \wt\sigma(3)=2, \,\wt\sigma(2)=3$ (note that these permutations are  really tested in step~2.(b)), we find 
\begin{align*}
\B_1=\begin{pmatrix}
1 & 1/10 & 1/3\\
0 & 9/10 & 1/3\\
0 & 0 & 1/3
\end{pmatrix}\quad \text{and}\quad 
\B_2=\begin{pmatrix}
1 & 1/10 & 1/3\\
0 & 17/30 & 0\\
0 & 1/3 & 2/3
\end{pmatrix}.
\end{align*}
As can be verified by  Theorem~4.2 of \cite{GK1}), the matrix $\B_1$ is the \MLCM\ of a \mSEM\ on the \DAG\ $\D_1$ depicted in \autoref{examV0ng}. 
Although $\sgn(B_2)$ is the reachability matrix of a \DAG, namely of the \DAG\ $\D_2$ from \autoref{examV0ng}, which is a necessary property of a matrix to be the \MLCM\ of a \mSEM\  according to \eqref{reach}, 
 it is no \MLCM\ of a \mSEM.
\eexam

\subsubsection*{Recursive max-weighted models}

Assume now that $\chi$ is the \TDM\ of a \RMWM. We modify and adapt \autoref{alg:Bconsistent} to obtain  a procedure which outputs the standardized \MLCM s of all \RMWM s  with \TDM\ $\chi$. Among the maximum $\chi$-cliques which we find in step~2.(a) of \autoref{alg:Bconsistent} are the initial node sets of the  \DAG s underlying the \RMWM s that have \TDM\ $\chi$. We learn from   \autoref{IdentifyMwm} and \autoref{Bstand+chi} that a maximum $\chi$-clique is such an  initial node set
iff the matrix $\B$ obtained by \autoref{BChiV0} is the \MLCM\ of a \RMWM. In that case, $\B$ is obviously the standardized \MLCM\ of a \RMWM\ with \TDM\ $\chi$. These observations lead to  the following procedure.

\balg[Modification of \autoref{alg:Bconsistent} for \RMWM s: find all $\B$  from $\chi$\label{alg:BconsistentMW}]

\begin{citemize}
\item[1.] Find all maximum $\chi$-cliques (cf. step~1. of \autoref{alg:Bconsistent}). 
\item[2.] For every maximum $\chi$-clique $W$,
\begin{citemize}
\item[(a)] check $\chi(i,j)\ge \sum_{k\in W} \chi(k,i)\wedge \chi(k,j)$ for all $i,j\in V\setminus W$;\\ 
 if not, then there is no  \RMWM\  with \TDM\ $\chi$ on a \DAG\ with initial nodes $W$; \\
else, 
\begin{citemize}
\item[i.] apply \autoref{BChiV0} with $V_{\text{\smaller$0$}}=W$; 
\item[ii.] check the following properties for the matrix $\B$ obtained in i.: 
\begin{citemize}
\item[-] $\sgn(\B)$ is the reachability matrix of a \DAG
\item[-] for all $i\in V$, $j\in\an(i)$, and $k\in\des(j)\cap\pa(i)$, $\b_{ji}=\frac{\b_{jk}\b_{ki}}{\b_{kk}}$
\end{citemize}
if not, 
then there is  no \RMWM\ with \TDM\ $\chi$ on a \DAG\ with initial nodes $W$;\\
else, $\B$ is the  standardized \MLCM\ of  a \RMWM\ with \TDM\ $\chi$. 
\end{citemize}
\end{citemize}
\end{citemize}
\ealg 

That the properties we verify for the matrix $\B$ in step~ii. are sufficient for $\B$ to be the \MLCM\ of a \RMWM\ can be verified by Corollary~4.3(a) of \cite{GK1}.

To conclude this section, we highlight  the essential steps of \autoref{alg:BconsistentMW} with an example.

\bexam[The class of \RMWM s is not closed under $\chi$-equivalence\label{exam:AlgChinc3}]\\
Consider the \TDM 
\vspace*{-1.35em}
\begin{multicols}{2}
\begin{minipage}{0.93\textwidth}
\begin{align*}
\chi=\begin{pmatrix}
1 & 0 & 0.2 & 0\\
0& 1 & 0.6 & 0.5\\
0.2 & 0.6& 1 & 0.5\\
0 &0.5& 0.5 & 1
\end{pmatrix}.
\end{align*}
\end{minipage}
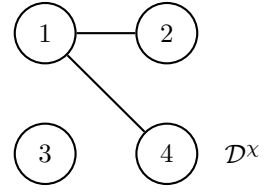
\begin{minipage}{0.5\textwidth}
\hspace*{10em}
\begin{tikzpicture}[-,every node/.style={circle,draw},line width=0.8pt, node distance=1.6cm,minimum size=0.8cm]
\node (n1)  {${1}$};
\node (n2) [right of=n1] {${2}$};
\node (n3) [below of=n1] {${3}$};
\node (n4) [right of=n3] {${4}$};
\node (n5)[draw=white,fill=white,right of=n4,node distance=1cm] {$\D^\chi$};
\foreach \from/\to in {n1/n2,n1/n4}
\draw[-,line width=0.8pt] (\from) -- (\to);
\end{tikzpicture}
\end{minipage}
\end{multicols} 
\vspace*{-0.3em}
\noindent 
We read from the complement   $\D^\chi$ of the $\chi$-graph that the sets $W_1=\{1,2\}$ and $W_2=\{1,4\}$ are the maximum $\chi$-cliques. Applying \autoref{BChiV0} with $V_{\text{\smaller$0$}}=W_1$ and $V_{\text{\smaller$0$}}=W_2$, we get the matrices 
\vspace*{-0.5em}
\begin{multicols}{3}
\begin{minipage}{0.18\textwidth}
\flushleft
\hspace*{-1em}
\begin{tikzpicture}[->,every node/.style={circle,draw},line width=0.8pt, node distance=1.6cm,minimum size=0.8cm,outer sep=1mm]
\node (n1)  {${1}$};
\node (n2) [right of=n1] {${2}$};
\node (n3) [below of=n1] {${3}$};
\node (n4) [right of=n3] {${4}$};
\node (n5)[draw=white,fill=white,left of=n3,node distance=0.88cm] {$\D_1$};
\foreach \from/\to in {n1/n3,n2/n3,n2/n4}
\draw[->, line width=0.8pt] (\from) -- (\to);
\end{tikzpicture}
\end{minipage}
\begin{minipage}{0.2\textwidth}
\begin{align*}
\hspace*{-5.5em}
\B_{1} =
\begin{pmatrix}
1	&	0&	0.2&	0\\
0	&	1   & 0.6	&0.5	\\
0	&0	&0.2	&0	\\
0	&	0 &  0	&0.5
\end{pmatrix}
\quad \text{and} \quad
\B_2=\begin{pmatrix}
1	&	0&	0.2&	0\\
0	&	0.5&	0.1&	0\\
0	&	0&	0.2&	0\\
0	&	0.5&	0.5&1
\end{pmatrix}.
\end{align*}
\end{minipage}
\begin{minipage}{0.28\textwidth}
\flushright
\begin{tikzpicture}[->,every node/.style={circle,draw},line width=0.8pt, node distance=1.6cm,minimum size=0.8cm,outer sep=1mm]
\node (n1)  {${1}$};
\node (n2) [right of=n1] {${4}$};
\node (n3) [below of=n1] {${3}$};
\node (n4) [right of=n3] {${2}$};
\node (n5)[draw=white,fill=white,right of=n4,node distance=1cm] {$\D_2$};
\foreach \from/\to in {n1/n3,n2/n3,n2/n4,n4/n3}
\draw[->, line width=0.8pt] (\from) -- (\to);
\end{tikzpicture}
\end{minipage}
\end{multicols}
\vspace*{-0.3em}
\noindent 
The matrix $\B_1$ is the \MLCM\ of a \RMWM\ on $\D_1$, whereas $\B_2$ is not the \MLCM\ of a \RMWM, but it is the  \MLCM\ of a \mSEM\ on $\D_2$. 
Therefore, all \RMWM s  with \TDM\ $\chi$ have the same standardized \MLCM\ $\B_1$, and $\D_1$ is their associated \DAG. Furthermore, all these models are $\chi$-equivalent to the \mSEM s with standardized \MLCM\ $\B_2$. 
\vspace{-\partopsep} 
\eexam

\section{$\chi$-equivalent recursive max-linear models and their \DAG s\label{s5}}

In this section we mainly present interrelations between \DAG s of $\chi$-equivalent \mSEM s.

One of the best known equivalence relations 
on the set of \DAG s  is certainly the Markov equivalence: two \DAG s are Markov equivalent  if they entail the same conditional independence relations through  the Markov property; for a characterization of such \DAG s, see e.g. \citet{Verma1991}. The associated \DAG\  of 
a recursive linear Gaussian structural equation model can be identified from the distribution
only up to a Markov equivalence class (under the assumption of faithfulness; 
see e.g. \cite{Spirtes2016}).  In the following example we discuss the relation between 
 $\chi$-equivalence of  \mSEM s and   Markov equivalence of their associated \DAG s. 

\bexam[The difference between $\chi$-equivalence of \mSEM s and Markov equivalence of their \DAG s]
\begin{citemize}
\item[(1)] Undirected graphs  underlying Markov equivalent \DAG s coincide.  \autoref{exam:AlgChinc3} clarifies  that this does not hold for \DAG s of $\chi$-equivalent \mSEM s. Such \DAG s are  therefore not necessarily Markov equivalent. 
\item[(2)] For the \TDC s of a \mSEM\ $\bfx$ on $\D_1$, which is always a \RMWM, we have by  \autoref{chimwm}(b) that $\chi(1,3) < \chi(1,2)\wedge\chi(2,3)$. Since $\D_2$ has initial node $2$,  by \autoref{le:Ws}(c) there cannot be a \mSEM\   that  is $\chi$-equivalent to $\bfx$ on  $\D_2$. Thus although the \DAG s $\D_1$ and $\D_2$ are Markov equivalent, there exist no $\chi$-equivalent \mSEM s on $\D_1$ and $\D_2$.
\item[(3)] As can be verified by \autoref{th:chartdm},   \mSEM s on the Markov equivalent \DAG s $\D_1$  and $\D_3$ are always $\chi$-equivalent. 
This shows that there can be $\chi$-equivalent \mSEM s on Markov equivalent \DAG s. 
\vspace*{-0.4em}
\end{citemize}
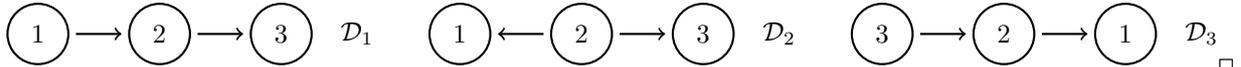
\begin{center}
\hspace*{0.5em}
\begin{tikzpicture}[->,every node/.style={circle,draw},line width=0.8pt, node distance=1.6cm,minimum size=0.8cm,outer sep=1mm]
\node (n1) {${1}$};
\node (n2) [right of=n1] {${2}$};
\node (n3) [right of=n2] {${3}$};
\node (n5)[draw=white,fill=white,,right of=n3,node distance=1cm] {$\D_1$};
\foreach \from/\to in {n1/n2,n2/n3}
\draw[->, line width=0.8pt] (\from) -- (\to);
\end{tikzpicture}
\hfill
\begin{tikzpicture}[->,every node/.style={circle,draw},line width=0.8pt, node distance=1.6cm,minimum size=0.8cm,outer sep=1mm]
\node (n1) {${1}$};
\node (n2) [right of=n1] {${2}$};
\node (n3) [right of=n2] {${3}$};
\node (n5)[draw=white,fill=white,,right of=n3,node distance=1cm] {$\D_2$};
\foreach \from/\to in {n2/n1,n2/n3}
\draw[->, line width=0.8pt] (\from) -- (\to);
\end{tikzpicture}
\hfill
\begin{tikzpicture}[->,every node/.style={circle,draw},line width=0.8pt, node distance=1.6cm,minimum size=0.8cm,outer sep=1mm]
\node (n1) {${3}$};
\node (n2) [right of=n1] {${2}$};
\node (n3) [right of=n2] {${1}$};
\node (n5)[draw=white,fill=white,right of=n3,node distance=1cm] {$\D_3$};
\foreach \from/\to in {n1/n2,n2/n3}
\draw[->, line width=0.8pt] (\from) -- (\to);
\end{tikzpicture}
\end{center}
\vspace{-10\partopsep} 
\eexam

\DAG s of $\chi$-equivalent \mSEM s have the same number of initial nodes, since  the initial node sets of such \DAG s are maximum $\chi$-cliques, which have the same cardinality by definition.  We learn from \autoref{BChiAlg} that if the standardized \MLCM s of two $\chi$-equivalent \mSEM s differ, then the causal orderings of their associated \DAG s must also differ. So for these two \DAG s there exist nodes $i,j\in V$ such that one \DAG\ has a path from $j$ to $i$ and the other has one from $i$ to $j$. We provide further properties of two \DAG s underlying $\chi$-equivalent \mSEM s.

\bpr\label{char:equiv} Let $\bfx$ and $\wt\bfx$ be $\chi$-equivalent \mSEM s on \DAG s $\D$ and $\wt\D$, respectively.  
We denote the initial nodes in $\D$ 
and $\wt\D$ by $V_{\text{\smaller$0$}}$ and $\wt V_{\text{\smaller$0$}}$, the ancestors of $i$ by $\an(i)$ and $\wt\an(i)$, and the descendants  of $i$  by $\des(i)$ and $\wt\des(i)$. 
\begin{citemize}
\item[(a)]  There is only one bijection $\varphi:V_{\text{\smaller$0$}} \to \wt V_{\text{\smaller$0$}}$ such that for every $j\in V_{\text{\smaller$0$}}$,   $\chi(j,\varphi(j))>0$ and $\chi(j,\wt j)=0$ for all $\wt j\in \wt V_{\text{\smaller$0$}}\setminus \{\varphi(j)\}$.  
\end{citemize}
Let $\varphi$ be the bijection from (a) and $j\in V_{\text{\smaller$0$}}$. 
\begin{citemize}
\item[(b)] We have  $\An(\varphi(j))\cap V_{\text{\smaller$0$}}=\wt \Des(\varphi(j))\cap V_{\text{\smaller$0$}}=\{j\}$. In particular, if $j\neq \varphi(j)$, then $\D$ has a path from $j$ to $\varphi(j)$, and $\wt\D$ has one from $\varphi(j)$ to $j$. 
\item[(c)] We have $\Des(j)=\wt\Des(\varphi(j))$. 
\item[(d)] For  $i\in V$, $\wt \An(i)\cap \wt V_{\text{\smaller$0$}}=\{ \varphi(j): j\in \An(i)\cap V_{\text{\smaller$0$}}  \}$. 
\end{citemize}
\vspace{-\partopsep} 
\epr
\bproof 
(a) is immediate by \autoref{le:Ws}(a), since $\wt V_{\text{\smaller$0$}}$ is a maximum $\chi$-clique.\\
(b) Since $\wt V_{\text{\smaller$0$}}$ is a maximum $\chi$-clique, according to \autoref{le:Ws}(b), $\An(\varphi(j))\cap V_{\text{\smaller$0$}}=\{j\}$. Note that for every $\wt j\in \wt V_{\text{\smaller$0$}}$, $\chi(\wt j,\varphi^{-1}(\wt j))>0$ and $\chi(\wt j,j)>0$ for all $j\in V_{\text{\smaller$0$}}\setminus \{\varphi^{-1}(\wt j)\}$,  where  $\varphi^{-1}:   \wt V_{\text{\smaller$0$}} \to V_{\text{\smaller$0$}}$ denotes the inverse of $\varphi$. As  $V_{\text{\smaller$0$}}$ is a maximum $\chi$-clique, we therefore have again by  \autoref{le:Ws}(b) that $\wt \Des(i)\cap V_{\text{\smaller$0$}}=\{\varphi^{-1}(i)\}$ with $i=\varphi(j)$, which is obviously equivalent  to $\wt \Des(\varphi(j))\cap V_{\text{\smaller$0$}}=\{j\}$.\\ 
(c)  Let $i\in \Des(j)$. By (b)  $j\in  \An(\varphi(j))\cap\An(i)$ and, consequently,  by \autoref{chiequiv} $\chi(\varphi(j),i)>0$. \autoref{lemW}(c) 
then  yields that $i\in \wt\Des(\varphi(j))$. Hence,  $\Des(j)\subseteq\wt\Des(\varphi(j))$. From this, by reversing the roles of $\D$ and $\wt\D$ and noting that  $\chi(\wt j,\varphi^{-1}(\wt j))>0$ for all $\wt j\in \wt V_{\text{\smaller$0$}}$, we observe that $ \wt\Des(\varphi(j))\subseteq \Des(j)$.  \\
(d) can be verified by (c).
\vspace{-\partopsep} 
\eproof

\subsection*{Recursive max-weighted models}

Now we consider $\chi$-equivalent \RMWM s and investigate their \DAG s.  Because of \autoref{RemMaxCl}, \autoref{BChiV0}, and \autoref{propBquer}(e), if a \TDM\ $\chi$ of a \RMWM\ has  one maximum $\chi$-clique $W$,  all \RMWM s with \TDM\ $\chi$ (the models are then $\chi$-equivalent by definition) have  the same standardized \MLCM\ and, hence, the same minimum \ML\ \DAG, which again has  initial nodes $W$. By \autoref{BChiV0} the initial nodes of \DAG s of $\chi$-equivalent \mSEM s with different standardized \MLCM s must also differ. We present further interrelationships between \DAG s of $\chi$-equivalent \RMWM s with regard to their initial nodes.

\bthe\label{char:equiv:mw} 
Let $\bfx$ and $\wt \bfx$ be $\chi$-equivalent \RMWM s on \DAG s $\D$ and $\wt\D$, respectively.  
We denote by $V_{\text{\smaller$0$}}$ and $\wt V_{\text{\smaller$0$}}$ the initial nodes  in   $\D$ and $\wt\D$ and by $V_{\!\text{\smaller$\infty$}}$ and $\wt V_{\!\text{\smaller$\infty$}}$ their terminal nodes.  Let $\varphi: V_{\text{\smaller$0$}}\to \wt V_{\text{\smaller$0$}}$ be the bijection from \autoref{char:equiv}(a) and $j\in V_{\text{\smaller$0$}}$ such that  $j\neq \varphi(j)$.  
\begin{citemize}
\item[(a)] We have $\varphi(j)\in V_{\!\text{\smaller$\infty$}}$. In particular, 
  $\wt V_{\text{\smaller$0$}} \subseteq (V_{\text{\smaller$0$}}\cap\wt  V_{\text{\smaller$0$}})\cup V_{\!\text{\smaller$\infty$}}$.
\item[(b)]  If $p=[j=k_0\to k_1\to\ldots\to k_{n-1}\to k_n= \varphi(j)]$ is a path in the transitive reduction $\D^\tr$ of $\D$, then $\wt p=[\varphi(j)=k_{n}\to k_{n-1}\to\ldots\to k_1\to k_0=j]$ is a path in the transitive reduction $\wt \D^\tr$ of $\wt\D$.
\end{citemize}
\vspace{-\partopsep} 
\ethe
\bproof 
We denote  by $\an(i)$ and $\wt \an(i)$  the ancestors of  $i$ in $\D$ and $\wt\D$ and by $\des(i)$ and $\wt \des(i)$  its  descendants.   \\
(a) Assume that $\varphi(j)\not\in V_{\!\text{\smaller$\infty$}}$. Consequently,  by \autoref{char:equiv}(b)  $\D$ has a path  from $j$ to some  $i\neq\varphi(j)$  passing through $\varphi(j)$. Replacing  $V_{\text{\smaller$0$}}$ by $\wt V_{\text{\smaller$0$}}$, we learn from the   the proof of \autoref{le:Ws}(c)  that $\chi(j,i)\ge \chi(\varphi(j),j) \wedge \chi(\varphi(j),i)$.  But this contradicts  \autoref{chimwm}(b). Hence, $\varphi(j)\in V_{\!\text{\smaller$\infty$}}$. \\
(b) Let $p$ be a path in $\D^\tr$.  To prove  that $\wt p$ is a path in $\wt \D^\tr$, because of the properties of $\wt\D^\tr$, it suffices to show that for  $\nu=0,\ldots,n-1$,  $k_{\nu+1}\in \wt \an(k_\nu)$ and $\wt \des(k_{\nu+1})\cap\wt \an(k_{\nu})\neq\emptyset$. Recalling from \autoref{char:equiv}(b) that $\An(\varphi(j))\cap V_{\text{\smaller$0$}}=\{j\}$, we observe that $\An(k_\nu)\cap \An(k_{\nu+1})\cap V_{\text{\smaller$0$}}=\{j\}$. We then obtain from  \autoref{char:equiv}(d)  that  $\wt \An(k_\nu)\cap \wt \An(k_{\nu+1})\cap \wt V_{\text{\smaller$0$}}=\{\varphi(j)\}$. By \autoref{chimwm}(b) we have $\chi(k_\nu,\varphi(j))= \chi(k_{\nu}, k_{\nu+1}) \chi(k_{\nu+1},\varphi(j))$. As $\wt \An(k_\nu)\cap \wt \An(k_{\nu+1})\cap \wt V_{\text{\smaller$0$}}=\{\varphi(j)\}$, using  \autoref{chimwm1} then proves that $k_{\nu+1}\in \wt \an(k_\nu)$. To show that  $\wt \des(k_{\nu+1})\cap\wt \an(k_{\nu})\neq\emptyset$, assume the converse. Let  $\ell\in\wt \des(k_{\nu+1})\cap\wt \an(k_{\nu})$.
By reversing the roles of $\D^\tr$ and $\wt\D^\tr$ and noting that for every $\wt j\in \wt V_{\text{\smaller$0$}}$, $\chi(\wt j,\varphi^{-1}(\wt j))>0$ and $\chi(\wt j,j)>0$ for all $j\in V_{\text{\smaller$0$}}\setminus \{\varphi^{-1}(\wt j)\}$,  where  $\varphi^{-1}:   \wt V_{\text{\smaller$0$}} \to V_{\text{\smaller$0$}}$ denotes the inverse of $\varphi$,
 we know from above that then $k_\nu\in\an(\ell)$ and $\ell\in\an(k_{\nu+1})$, i.e., $\des(k_\nu)\cap\an(k_{\nu+1})\neq \emptyset$. But this is in contradiction to the fact that $p$ is a path in $\D^\tr$. Hence, $\wt\D^\tr$ must contain $\wt p$. 
\eproof

In the next example we use \autoref{char:equiv:mw}  to find \RMWM s  that are $\chi$-equivalent to a given one. 
 
\bexam[Continuation of \autoref{exam:lambda}: find $\chi$-equivalent \RMWM s\label{exam:lambda1}]\\
By \autoref{chiequiv}  the sets $\{1\},\ldots, \{99\}$ are the maximum $\chi$-cliques. Since $99$ is the only terminal node in 
 $\D$, 
 it may be the only initial node of a \DAG\ that underlies  a potential \RMWM\ with the  same \TDM\ $\chi$ as $\bfx$ and  differs from $\D$. 
 Thus 
 the \DAG\ 
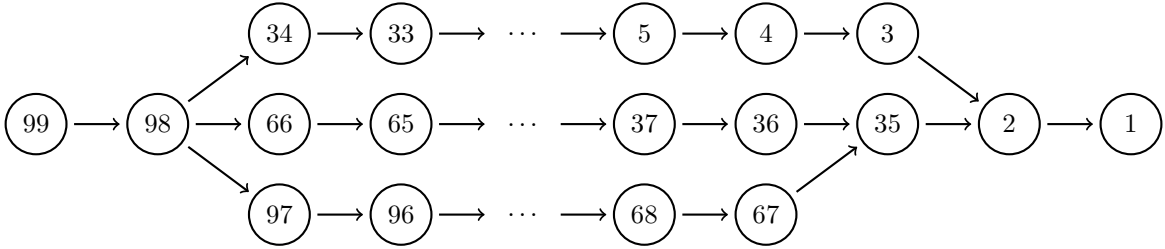
\vspace*{-0.25em}
\begin{center}
\begin{tikzpicture}[->,every node/.style={circle,draw},line width=0.8pt, node distance=1.6cm,minimum size=0.8cm,outer sep=1mm]
\node (1)  {$99$};
\node (2) [right of=1]  {$98$};  
\node (35) [right of=2]  {$66$};  
\node (36) [right of=35]  {$65$};  
\node (37) [draw=white,right of =36]{$\ldots$};
\node (38) [right of =37]{$37$};
\node (65) [right of =38]{$36$};
\node (66) [right of =65]{$35$};
  \node (98) [right of =66]{$2$};
  \node (99) [right of =98]{$1$};  
  \node (3) [above of=35,node distance=1.2cm]  {$34$};  
    \node (4) [above of=36,node distance=1.2cm]  {$33$};      
        \node (5) [draw=white,above of=37,node distance=1.2cm]  {$\ldots$};      
    \node (6) [above of =38,node distance=1.2cm]{$5$};
\node (33) [above of =65,node distance=1.2cm]{$4$};  
\node (34) [above of =66,node distance=1.2cm]{$3$};  
\node (67) [below of =35,node distance=1.2cm]{$97$};  
\node (68) [right of =67]{$96$};
\node (69) [draw=white,right of =68]{$\ldots$};
\node (96) [right of =69]{$68$};  
\node (97) [below of =65,node distance=1.2cm]{$67$};  
    \foreach \from/\to in {1/2,2/35,35/36,36/37,37/38,38/65,65/66,66/98,98/99,2/3,3/4,4/5,5/6,6/33,33/34,34/98,2/67,67/68,68/69,69/96,96/97,97/66}
  \draw (\from) -- (\to);
\end{tikzpicture}
\end{center}
\vspace*{-0.3em}
is  the transitive reduction $\wt\D^\tr$ of such a \DAG. 
 To verify  the existence of a \RMWM\ with \TDM\ $\chi$  on a \DAG\ whose transitive reduction is $\wt\D^\tr$,  we may compute the matrix $\B$ from \eqref{eq:BChi1} and check then whether it is the \MLCM\ of a \RMWM. 
\eexam

We conclude this section with an example investigating whether a \RMWM\  on a known \DAG\ is $\chi$-equivalent to a \RMWM\ on another given  \DAG. 

\bexam[The existence of  $\chi$-equivalent \RMWM s on given \DAG s\label{exam:AlgChinc1}]\\
 We consider a \RMWM\  $\bfx$  with \TDM\ $\chi$ on $\D_1$ and clarify when $\bfx$ is $\chi$-equivalent to a \RMWM\ on $\D_2$.
Note that all \mSEM s on $\D_1$ and on $\D_2$ are max-weighted. By \autoref{th:chartdm} we find  
\begin{align*}
\chi(1,2)= 0, \,\,\, \chi(1,4)=0,\,\,\, \chi(1,3)>0, \,\,\, 1-\chi(1,3)-\chi(2,3)>0, \,\,\, & 1-\chi(2,4)>0,   \,\,\,  \chi(3,4)=\chi(2,3)\wedge\chi(2,4)>0
\end{align*}
and also that $\chi$ is the \TDM\ of a \RMWM\ on $\D_2$ iff 
\begin{align*}
\chi(1,2)= 0, \,\,\, \chi(1,4)=0,\,\,\, \chi(1,3)>0, \,\,\, 1-\chi(1,3)-\chi(3,4)>0, \,\,\, & 1-\chi(2,4)>0,   \,\,\, \chi(2,3)=\chi(2,4)\wedge\chi(3,4)>0.
\end{align*}
This implies that $\bfx$ is $\chi$-equivalent to a \RMWM\ on $\D_2$ iff $\chi(2,3)=\chi(3,4)$.

As shown in \autoref{exam:AlgChinc3} the matrix $\chi$ given therein
is the \TDM\ of a \RMWM\ on $\D_1$. As $\chi(2,3)=0.6 \neq\chi(3,4)=0.5$, such a model cannot be $\chi$-equivalent to a \RMWM\ on $\D_2$. Of course, we already know this from \autoref{exam:AlgChinc3}. 
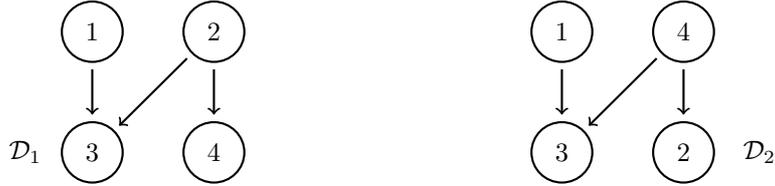
\vspace*{-0.75em}
\begin{center}
\begin{tikzpicture}[->,every node/.style={circle,draw},line width=0.8pt, node distance=1.6cm,minimum size=0.8cm,outer sep=1mm]
\node (n1)  {${1}$};
\node (n2) [right of=n1] {${2}$};
\node (n3) [below of=n1] {${3}$};
\node (n4) [right of=n3] {${4}$};
\node (n5)[draw=white,fill=white,left of=n3,node distance=0.88cm] {$\D_1$};
\foreach \from/\to in {n1/n3,n2/n3,n2/n4}
\draw[->, line width=0.8pt] (\from) -- (\to);
\end{tikzpicture}
\hspace*{10em}
\begin{tikzpicture}[->,every node/.style={circle,draw},line width=0.8pt, node distance=1.6cm,minimum size=0.8cm,outer sep=1mm]g
\node (n1)  {${1}$};
\node (n2) [right of=n1] {${4}$};
\node (n3) [below of=n1] {${3}$};
\node (n4) [right of=n3] {${2}$};
\node (n5)[draw=white,fill=white,right of=n4,node distance=1cm] {$\D_2$};
\foreach \from/\to in {n1/n3,n2/n3,n2/n4}
\draw[->, line width=0.8pt] (\from) -- (\to);
\end{tikzpicture}
\end{center}
\vspace{-10\partopsep} 
\eexam

\section{Conclusion\label{s6}}

A \mSEM\ is not restricted to heavy-tailed noise variables, but is defined in \cite{GK1} for independent noise variables with support $\R_+$.
Only, if the noise variables are heavy-tailed, the \TDM\ is meaningful (not identical to 0) for modeling the dependence structure in a \mSEM.

In this heavy-tailed setting, we considered the problem of identifying a \mSEM\ $\bfx$ on a \DAG\ $\D$ from its \TDM\ $\chi$. Simply because of the symmetry of $\chi$, the identifiability of $\bfx$ is not possible in general. \mSEM s  with arbitrary index of regular variation and     \MLCM\  whose column sums are also arbitrary have \TDM\ $\chi$. As our focus was on the causal structure of $\bfx$ represented by $\D$, we concentrated on the standardized model, where  the index of regular variation is one and the  columns of its \MLCM\ $\B$ add up to one. We showed that $\B$ 
can be recovered from $\chi$ and some additional information on $\D$  such as the full reachability relation or only a causal ordering. In these situations we can also determine 
the minimum \ML\ \DAG\ $\D^B$ of $\bfx$, the  smallest \DAG\ which represents the recursive max-linear dependence structure of $\bfx$. We  developed an algorithm which outputs the standardized \MLCM s of all \mSEM s having \TDM\ $\chi$. Moreover, we found the \RMWM s as a relevant subclass of \mSEM s. 
The simple structure of their \TDM s allows for identifiability of $\B$ and $\D^B$ from $\chi$ and the initial nodes of $\D$. This led to a  simpler approach to find the standardized \MLCM s of all \RMWM s with \TDM\ $\chi$.

Future work will focus on statistical properties of \mSEM s.

\section*{Acknowledgements}
We thank Steffen Lauritzen and Jonas Peters for interesting discussions and Zhongwei Zhang for his careful reading of our manuscript. MO and NG thank the  International Graduate School of Science and Engineering (IGSSE) of the Technical University of Munich for support.

\section*{References}


\appendix

\section{Appendix}

\subsection{Properties of the standardized max-linear coefficient matrix of a recursive max-linear model\label{A1}}

We summarize some properties of the standardized \MLCM\ $\B$ defined in  \eqref{Bquer}, which are used throughout the paper. 

\ble  \label{propBquer}
Let $\bfx$ be a \mSEM\ on a  \DAG\ $\D$ with \MLCM\ $B$ and standardized \MLCM\ $\B$. 
\begin{citemize}
\item[(a)] We have $\sgn(\B)=\sgn(B)$. 
\item[(b)] For $i\in V$, $\sum_{k\in\An(i)}\b_{k i}=\sum_{k=1}^d\b_{ki}=1$. 
\item[(c)] The matrix $\B$ is the  \MLCM\  of a \mSEM\  on $\D$. 
\item[(d)] For $i\in V$, $k \in\an(i)$, and $j\in \an(k)$, $\b_{j i}\ge \frac{\b_{jk}\b_{ki}}{\b_{kk}}$ with equality iff there is a max-weighted path from $j$ to $i$ passing through $k$.  
\item[(e)] The minimum \ML\ \DAG s $\D^B$ and $\D^\B$ coincide.
\item[(f)] For distinct $i,j\in V$, $\b_{jj}>\b_{ji}$.
\end{citemize}
\vspace{-\partopsep} 
\ele
\bproof
(a) and (b) are immediate consequences of the definition of $\B$ and \eqref{reach}.\\
(c) can be verified by Theorem~4.2 of  \cite{GK1}.\\
(d) The inequality follows from (c) and Corollary~3.12  of \cite{GK1} and the rest of the statement from Theorem~3.10(a)  of  \cite{GK1} and by observing that $\b_{ji}=\frac{\b_{jk}\b_{ki}}{\b_{kk}}$ iff $b_{ji}=\frac{b_{jk} b_{ki}}{b_{kk}}$. \\
(e) is a consequence of Theorem 5.3 of \cite{GK1} and the definition of $\B$. \\ 
(f) For $j\in V\setminus\An(i)$ we have  immediately by (a) that  $\b_{ji}=0 < \b_{jj}$. For $j\in \An(i)$ we obtain by parts (b) and (d),
\begin{align*}
1= \sum_{k\in \An(j)}\b_{k i}+ \sum_{k\in \An(i)\setminus \An(j)} \b_{k i}
\ge \frac{\b_{ji}}{\b_{jj}} \sum_{k\in \An(j)} \b_{k j} +  \sum_{k\in \An(i)\setminus \An(j)} \b_{k i}
= \frac{\b_{ji}}{\b_{jj}} + \sum_{k\in \An(i)\setminus \An(j)} \b_{k i}.
\end{align*}
Since $\An(i)\setminus \An(j)\neq \emptyset$ 
and  $\b_{k i}>0$ for all $k \in \An(i)\setminus \An(j)$, 
we find $1> \frac{\b_{ji}}{\b_{jj}}$, equivalently $\b_{jj}>\b_{ji}$.
\vspace{-\partopsep} 
\eproof

\subsection{Derivation of the tail dependence matrix of a recursive max-linear model\label{A2}}

We first prove \eqref{maxdom} and specify $G$ and its univariate and bivariate marginal distributions. 

\bpr\label{prop2.1}
Let $\bfx$ be a \mSEM\ on a  \DAG\ $\D$ with \MLCM\ $B$. Then $\bfx\in\MDA(G)$ with
\begin{align*} 
G(\textbf{x})=\exp\Big\{- \sum_{j=1}^d \bigvee_{i\in\Des(j)} \Big(\frac{b_{ji}}{x_{i}}\Big)^\al\Big\}, \quad \bfsx=(x_1,\ldots,x_d)\in\R_+^d. 
\end{align*}
Let  $\bs{M}=(M_1,\ldots,M_d)$ be a random vector with distribution function $G$. Then for $i,j \in V$  the distribution functions of $M_i$ and $(M_i,M_j)$ are given by
\begin{align*}
G_i(x_i) = \exp\Big\{-x_i^{-\al}\sum_{j\in\An(i)} b_{ji}^\al\Big\} \quad \text{and} \quad G_{ij}(x_i,x_j) = \exp\Big\{-\sum_{k\in\An(i)\cup\An(j)} \Big(\frac{b_{ki}}{x_i}\Big)^\al\vee \Big(\frac{b_{kj}}{x_j}\Big)^\al\Big\}. 
\end{align*}
\epr
\bproof 
As $Z\in\text{RV}(\al)$, 
there exists a normalizing sequence $a_n\in\R_+$  such that for every $x\in\R_+$,
\begin{align}\label{noiseRegVar}
\lim_{n\to\infty}F^n_{Z}(a_nx)=\Phi_\al (x)
\end{align} 
\citep[e.g.][Proposition~1.11]{Resnick1987}. Using \eqref{ml-noise}, 
 the independence of the noise variables, and \eqref{noiseRegVar}, we obtain for $\bfsx\in\R_+^d$,
\begin{align*}
\big[\P\big(\bfx\le a_n\bfsx\big)\big]^n&=\big[ \P\big(\bigvee_{j\in\An(i)}b_{ji}Z_j \le a_n x_i,\, i\in V\big)\big]^n \nonumber\\
&=\big[ \P\big(Z_j \le a_n\bigwedge_{i \in \Des(j)} \frac{x_i}{b_{ji}},\, j\in V\big)\big]^n \nonumber\\
&= \prod_{j=1}^d F^n_Z\big(a_n \bigwedge_{i \in \Des(j)} \frac{x_i}{b_{ji}}  \big) \nonumber \\
& \xrightarrow[n\to\infty]{}  \, \prod_{j=1}^d \Phi_\al\big( \bigwedge_{i \in \Des(j)} \frac{x_i}{b_{ji}} \big)
= G(\bfsx).
\end{align*}
This proves that $\bfx\in\MDA(G)$ 
(cf. Eq.~\eqref{maxdom}). 
Finally, the distribution functions of $M_i$ and $(M_{i},M_j)$  
are obtained by letting all other components of $\bfsx$ in $G$ tend to $\infty$ and recalling \eqref{reach}. 
\eproof 

\bproof[Proof of \eqref{tdc}] 
For every $k\in V$  we have $n(1-F_k(a_{k,n}))\to 1$ as $n\to \infty$  with $a_{k,n}:=F_k^{\leftarrow}\big(1-\frac 1n\big)=\big(\frac{1}{1-F_k}\big)^{\leftarrow}(n)$. 
Thus, 
\begin{align*}
\chi(i,j)& = \lim_{n\to \infty} \frac{\P(X_i > a_{i,n}, X_j > a_{j,n})}{1-F_j(a_{j,n})}\\
&=  \lim_{n\to \infty} n [1-F_i(a_{i,n})+1 - F_j(a_{j,n})-1+ \P(X_i\le a_{i,n}, X_j\le a_{j,n}) ]\\
&= 2 - \lim_{n\to\infty} n[1-\P(X_i\le a_{i,n}, X_j\le a_{j,n})]. 
\end{align*}
By Proposition~5.10(b), whose conditions are satisfied according to
 \autoref{prop2.1}, and Eq.~(5.38) of \citet{Resnick1987}, we find 
\begin{align*}
\chi(i,j) &=  2+ \log G_{ij} (( - 1/\log G_i)^{\leftarrow}(1),   ( - 1/\log G_j)^{\leftarrow}(1) ),
\end{align*}
where $(- 1/\log G_i)^{\leftarrow}$ and $(- 1/\log G_j)^{\leftarrow}$ denote the generalized inverses of the functions $- 1/\log G_i$ and $- 1/\log G_j$. 
With the representations  for $G_i$, $G_j$, and $G_{ij}$ from \autoref{prop2.1}, we then obtain by a simple calculation
\begin{align*}
\chi(i,j)=2-\sum_{k\in\An(i)\cup\An(j)} \b_{ki}\vee\b_{kj}.
\end{align*}
Finally, using \autoref{propBquer}(b), (a) yields 
  \begin{align*}
\chi(i,j)&=\sum_{k\in\An(i)\cup\An(j)}\b_{ki}+\sum_{k\in\An(i)\cup\An(j)}\b_{kj}-\sum_{k\in\An(i)\cup\An(j)} \b_{ki} \vee \b_{kj}\\
& =\sum_{k \in \An(i) \cup \An(j)} \b_{ki} \wedge \b_{kj}=\sum_{k \in \An(i) \cap \An(j)} \b_{ki} \wedge \b_{kj}. 
 \end{align*}
We learn from this proof that $\bfx$ and  the limit vector $\bs{M}$ from \eqref{maxdom} have the same \TDM, since $\bs{M}\in\MDA(G)$. 
\eproof

\end{document}